\documentclass[5p,twocolumn,10pt,times]{elsarticle}
\usepackage{amsmath}
\usepackage{hyperref}
\usepackage[switch]{lineno}
\graphicspath{{./images/}}
\usepackage{my}

\begin{document}
\baselineskip11pt

\begin{frontmatter}

\title{BRepMAE: Self-Supervised Masked BRep Autoencoders for Machining Feature Recognition}

\author[one]{Can Yao} 
\author[zero]{Kang Wu} 
\author[zero]{Zuheng Zheng} 
\author[one]{Siyuan Xing} 
\author[zero]{Xiao-Ming Fu} 

\address[one]{School of Artificial Intelligence and Data Science, University of Science and Technology of China, Hefei, 230026, People's Republic of China}
\address[zero]{School of Mathematical Sciences, University of Science and Technology of China, Hefei, 230026, People's Republic of China}



\begin{abstract}
We propose a masked self-supervised learning framework, called \emph{BRepMAE}, for automatically extracting a valuable representation of the input computer-aided design (CAD) model to recognize its machining features.
Representation learning is conducted on a large-scale, unlabeled CAD model dataset using the geometric Attributed Adjacency Graph (gAAG) representation, derived from the boundary representation (BRep).
The self-supervised network is a masked graph autoencoder (MAE) that focuses on reconstructing geometries and attributes of BRep facets, rather than graph structures.
After pre-training, we fine-tune a network that contains both the encoder and a task-specific classification network for machining feature recognition (MFR).
In the experiments, our fine-tuned network achieves high recognition rates with only a small amount of data (e.g., 0.1\% of the training data), significantly enhancing its practicality in real-world (or private) scenarios where only limited data is available.
Compared with other MFR methods, our fine-tuned network achieves a significant improvement in recognition rate with the same amount of training data, especially when the number of training samples is limited.
\end{abstract}


\end{frontmatter}


\section{Introduction} \label{sec:intro}
Machining feature recognition (MFR) is a technique that automatically extracts geometric shapes with specific machining semantics, such as holes, slots, bosses, and chamfers, from the 3D computer-aided design (CAD) model of the part.
Its core is to convert the geometric elements in the design model into understandable machining units (such as drilling and milling contours) in the manufacturing process, providing basic data for CNC programming~\cite{zhang2022machining}.
Thus, it is usually regarded as the fundamental step in automatic machining, being a link in CAD and computer-aided manufacturing (CAM) integration and transmitting design intent to the manufacturing process~\cite{ding2021mbd, siemiatkowski2021planning, wu2021fast, zhang2024novel, ji2016reachability, liu2021intelligent, givehchi2015generic}.

The MFR consists mainly of two tasks: (1) feature segmentation and (2) feature type identification.
The former segments the part into different geometric regions that constitute machining features.
The latter classifies these regions according to semantics.
Although the goal of MFR is clear, achieving accurate and effective results in practice remains challenging due to the complexity of real-world situations, such as the wide variety of feature types, the presence of different variants of the same feature, and the fusion of multiple features.

There are many existing MFR methods, which are often classified into two categories: (1) rule-based and (2) learning-based methods~\cite{shi2020critical, shi2020novel}.
The rule-based methods~\cite{zehtaban2016automated, venuvinod1995graph, parvaz2012multi, kim2014stepwise, xu2022automatic} require the manual definition of discriminative rules, which are not only time-consuming and labor-intensive but also lack generalization ability, rendering them ineffective in complex situations.
In contrast, learning-based methods~\cite{yao2023machining, ning2023part, shi2020intersecting, cao2020graph, zhang2022machining} can automatically learn classification rules from datasets, reducing labor consumption to some extent.

Most learning-based methods are supervised, requiring a manually labeled dataset and limiting their generalization to new datasets~\cite{chen2020big, ericsson2021well}.
To overcome these shortcomings, unsupervised (also known as self-supervised) methods~\cite{lou2023brep} and domain adaptation techniques~\cite{zhang2024brepmfr, zheng2025sfrgnn} have been proposed.
%
%
%
The self-supervised method~\cite{lou2023brep} trains a GNN tokenizer to generate discrete entity labels for BRep; however, due to the large number of similar edges and faces in BRep, it is prone to codebook collapse, limiting its performance on MFR. 
%
%
The domain adaptation methods~\cite{zhang2024brepmfr, zheng2025sfrgnn} rely on extensive, high-quality labeled data from the source domain and require the discriminator to be trained on a large amount of target-domain data, thereby reducing their practicality.
In this paper, we propose a self-supervised masked boundary representation (BRep) autoencoder, referred to as \emph{BRepMAE}, for MFR. 
This method first trains an autoencoder on a large-scale, unlabeled CAD dataset, then fine-tunes the network with limited labeled data for MFR.
During both training phases, including pre-training and fine-tuning, we utilize the geometric Attributed Adjacency Graph (gAAG) representation~\cite{wu2024aagnet}, which contains the geometric, attributed, and topological information of the CAD model.
To train the masked autoencoder, we randomly mask the geometries and attributes of most nodes in gAAG for the encoder and then recover the masked geometries and attributes using the decoder without changing the graph topology, inspired by GraphMAE~\cite{hou2022graphmae}.
After finishing training the autoencoder, we connect a multilayer perceptron (MLP) as a task-specific classification network behind the encoder to form a new network, which is fine-tuned using limited labeled MFR data.
Although several MAE models designed for 3D data have been proposed~\cite{liang2022meshmae, zhang2024point} and have demonstrated strong performance on downstream tasks that require general representation, such as feature recognition and object classification, no MAE model for BReps has been studied.
Therefore, the main contributions of this paper are as follows:
\begin{enumerate}
    \item \blu{We are the first to introduce MAE into B-Rep with a novel multi-branch reconstruction objective for self-supervised representation learning, enabling a generic encoder to capture both geometries and topologies.}
    \item \blu{The pre-trained representations remain effective under extremely low-label settings (e.g., 0.1\%-1\%), significantly reducing the reliance on labeled data and better matching practical industrial scenarios where CAD annotations are costly and rare.}
\end{enumerate}
%
%
%
%
In the experiments, we compare our fine-tuned network with three other MFR methods, including two supervised learning methods~\cite{wu2024aagnet, zhang2024brepmfr} and one self-supervised learning method~\cite{lou2023brep}, on various amounts of training data.
The results demonstrate that our fine-tuned network achieves high recognition accuracy with a limited amount of training data and consistently performs best in nearly all experiments, indicating the effectiveness and generalizability of our methods.
We also conduct a thorough ablation study of the network's components to validate their necessity.
%
%

\begin{figure*}[t]
 \centering
		\begin{overpic}[width=0.99\linewidth]{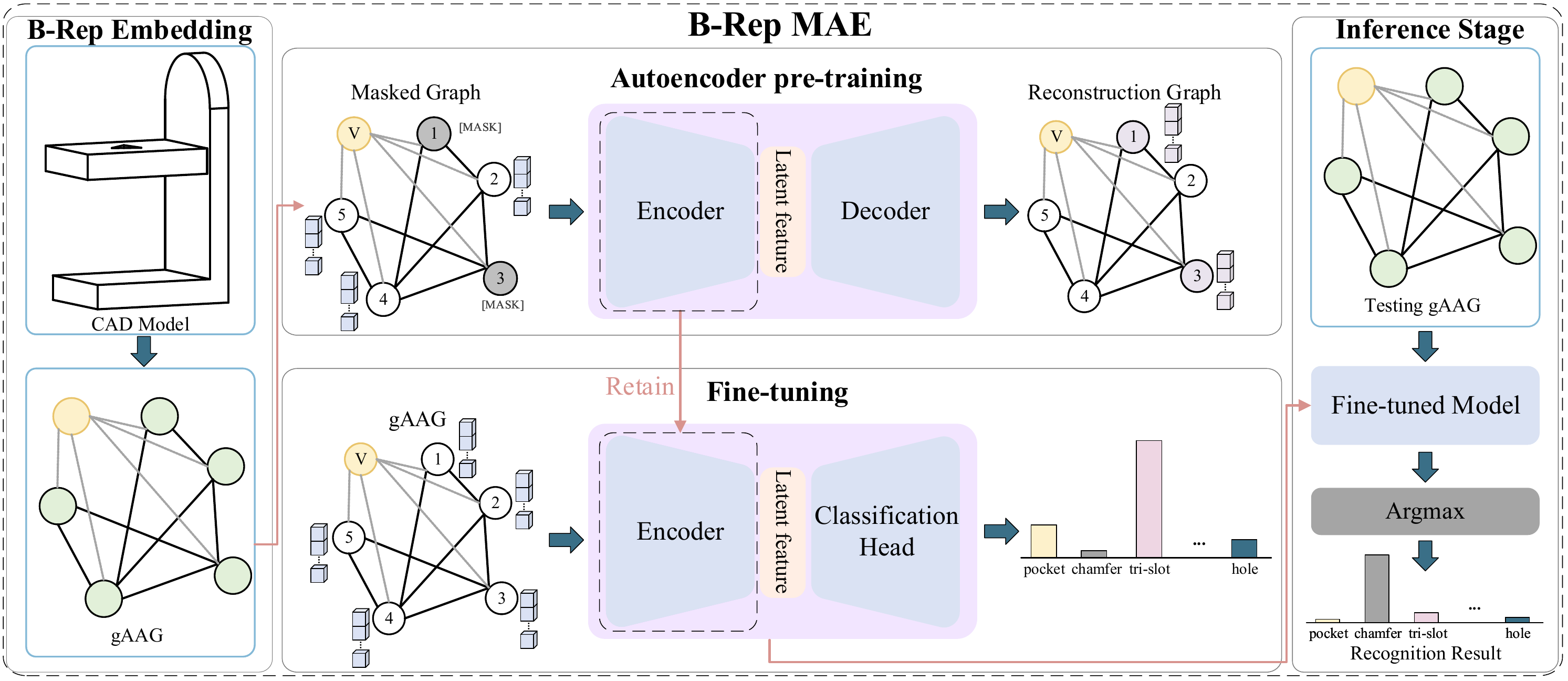}
			{
		    }
		\end{overpic}
		 \caption{
         Given a BRep, we first convert it into a gAAG, where graph nodes represent BRep facets and graph edges represent BRep edges.
         Then, we embed the graph node and edge features into 256-dimensional spaces and pre-train a masked autoencoder to learn the BRep representation.
         Finally, the fine-tuning stage retains the encoder of our autoencoder and replaces the decoder with a classification head to train this fine-tuned model on various amounts of labeled data for MFR.
         %
			 }
		 \label{fig:pipeline}
\end{figure*}

\section{Related work} \label{sec:related}
\paragraph{Rule-based MFR}
Rule-based methods rely on manually defined rules by professionals, which require specialized knowledge and experience.
Generally speaking, they can be roughly divided into five categories: logic rule-based methods~\cite{zehtaban2016automated, pratama2020rule, liu2018machining}, graph-based methods~\cite{ding2021mbd, venuvinod1995graph, cai2018freeform, yan2023manufacturing}, hint-based methods~\cite{han1998hint, parvaz2012multi, li2015hint, li2018local}, volume decomposition methods~\cite{kim2014stepwise, kataraki2017auto, kwon2019b}, and hybrid methods~\cite{xu2022automatic, guo2021hybrid, wang2023hybrid}.
Thus, they are often both labor-intensive and limited in generalization, often proving inadequate in complex scenarios.

\paragraph{Learning-based MFR}
The learning-based approach leverages neural networks' learning ability to extract useful information from datasets, automatically generating recognition feature rules. 
This method does not require manually defined rules, thereby reducing the consumption of human resources.
Since existing deep learning frameworks cannot accept BRep models as direct inputs~\cite{cao2020graph}, it is necessary to preprocess BRep models to make them compatible with these frameworks.

Point clouds, a common data structure, are used to represent BRep models~\cite {zhang2022machining, yao2023machining, lei2022mfpointnet, zhang2024point}.
%
However, point clouds lack a topological structure, which can lead to the loss of some information, and these methods are highly sensitive to the quality of the point clouds. 
%
Another common approach is to convert BRep models into multi-view images~\cite{shi2020novel, shi2020intersecting, shi2022highly}.
%
Then, techniques such as 2D CNNs are used to recognize machining features in images and extract them from parts.
However, they rely on the choice of views, which may lead to missing features in complex parts.
Besides, there are the voxel-based method~\cite{zhang2018featurenet}, which incurs excessive computational overhead, and the step-file-based method~\cite{miles2023approaching}, which must address the challenging task of extracting practical information from step files, rendering them unsuitable for MFR. 

With the development of graph neural networks, representing BRep models as graphs has begun. 
%
%
Compared with other representations, the graph shares the same topology and geometry information as the BRep model and has fewer elements than discrete structures~\cite{cao2020graph, colligan2022hierarchical}.
%
%
Thus, developing learning algorithms using graph representations for MFR has become popular~\cite{zhang2024novel, shi2020manufacturing, wu2024aagnet, ning2023part, wang2023hybrid1, zhang2022intelligent, liu2022supervised, zhang2024brepmfr, zheng2025sfrgnn, colligan2022hierarchical}.
\blu{For example, Hierarchical CADNet~\cite{colligan2022hierarchical} learns directly from B-Rep structures and demonstrates strong performance on MFR.}

%
%

\paragraph{Self-supervised learning for CAD models}
Self-supervised learning is widely applied across diverse domains because it can learn meaningful representations without relying on labeled data~\cite{liu2021self}.
In manufacturing, due to the lack of labeled CAD model data, self-supervised learning is introduced to reduce the data requirements.
These self-supervised learning methods are proposed for CAD retrieval~\cite{quan2024self}, representation learning~\cite{jones2023self, jung2024contrastcad, lou2023brep, ma2023multicad}, and CAD generation~\cite{xu2024brepgen, li2025dtgbrepgen, liu2025hola}.
\blu{Masked autoencoders have shown strong representation learning ability in vision and point clouds~\cite{he2022masked,pang2022masked,yu2022point}, motivating MAE-style pretraining for structured CAD data.}
However, although self-supervised learning has been widely applied in manufacturing, there are still gaps, such as the MFR problems associated with CAD models.
\blu{Thus, we develop an MAE-style self-supervised learning framework for extracting manufacturing features from BRep models, thereby unleashing its vast potential for widespread applications in this field.}


\section{Method} \label{sec:method}
\subsection{Overview of the proposed method}
Our masked self-supervised learning framework, called \emph{BRepMAE}, consists of three stages: data preprocessing, pre-training, and fine-tuning.
In the data preprocessing stage, we extract BRep represented as gAAGs, where gAAG nodes represent BRep faces, and edges represent BRep edges (Sec.~\ref{sec:data-preprocessing}).
Subsequently, in the pre-training stage, the nodes of the gAAG are randomly masked, i.e., the features of the masked nodes are replaced with learnable mask tokens.
An encoder-decoder architecture is then applied to reconstruct the masked feature, allowing to learn geometric and topological representations (Sec.~\ref{sec:pre-training}).
%
%
%
Finally, the fine-tuning stage retains the pre-trained encoder while replacing the decoder with task-specific heads for downstream tasks with labeled data, such as solid classification and surface segmentation (Sec.~\ref{sec:fine-tuning}). 
%
%
The overall pipeline of BRepMAE is shown in Fig.~\ref{fig:pipeline}.

\subsection{Data preprocessing}\label{sec:data-preprocessing}
%
%
Inspired by AAGNet~\cite{wu2024aagnet}, we extract and encode the geometry information and attributes of the BRep faces as the node feature values of the graph, and those of the BRep edges as the edge feature values of the graph.

\subsubsection{Geometric Information}
\begin{figure}[t]
 \centering
		\begin{overpic}[width=0.99\linewidth]{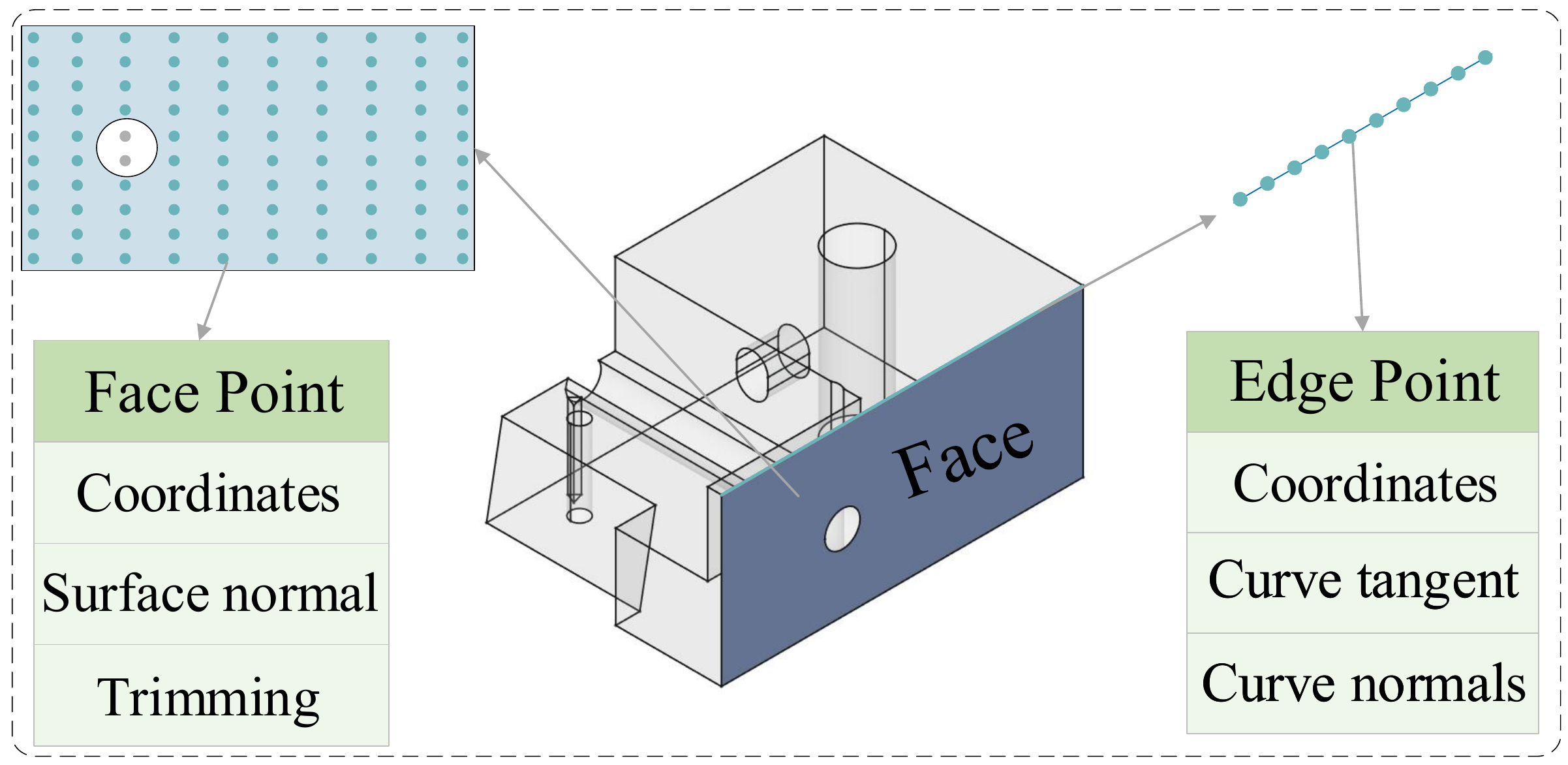}
			{
		    }
		\end{overpic}
		 \caption{
         For each face of a BRep model, we perform UV parameterization to discretize it into regular $10\times 10$ grid points.
         We extract the coordinate, surface normal, and trimming state of each grid point to form the face geometric information.
         Similarly, we uniformly sample 10 points in the curve parameter domain for each edge and extract the coordinates, tangent vectors, and normals of each point from both adjacent faces as the edge's geometric information.
			 }
		 \label{fig:geometry-info}
\end{figure}

We adopt UV parameterization to discretize B-Rep geometric entities into consistent representations~\cite{jayaraman2021uv}. 
\blu{Before discretization, we scale each input CAD model into a unit box to enhance training efficiency.}
The geometry of faces is discretized into regular $10 \times 10$ grids within UV parameter domains (see Fig.~\ref{fig:geometry-info} left). 
At each grid point, we extract 3D coordinates $(X^s, Y^s, Z^s)$, surface normals $(N_x^s, N_y^s, N_z^s)$, and the relationship between points and trimming surfaces $T^s$.
$T^s=0$ means this point is in the trimmed area; otherwise, $T^s=1$.
These features are combined into 7-dimensional vectors $(X^s, Y^s, Z^s, N^s_x, N^s_y, N^s_z, T^s)$, forming face geometric representation $f_{\text{geom}} \in \mathbb{R}^{7 \times 10 \times 10}$. 
Similarly, the geometry of curves is discretized using 10 uniformly sampled points along the curve parameter domains (see Fig.~\ref{fig:geometry-info} right). 
At each sample point, we extract 3D coordinates $(X^c, Y^c, Z^c)$, curve tangents $(T^c_x, T^c_y, T^c_z)$, and normal vectors from both adjacent faces, $(N^c_{l,x}, N^c_{l,y}, N^c_{l,z})$ and $(N^c_{r,x}, N^c_{r,y}, N^c_{r,z})$. 
These features are combined into 12-dimensional vectors $(X^c, Y^c, Z^c, T^c_x, T^c_y, T^c_z, N^c_{l,x}, N^c_{l,y}, N^c_{l,z}, N^c_{r,x}, N^c_{r,y}, N^c_{r,z})$, forming edge geometric representation $e_{\text{geom}} \in \mathbb{R}^{12 \times 10}$.
\blu{In our implementation, UV samples are generated on the face parameter domain, and the trimming state $T^s$ is obtained by querying the CAD kernel to test whether each $(u,v)$ sample lies inside the visible region of the face.}
\blu{Unless otherwise specified, we use a default UV-grid resolution of $10\times 10$ for faces and 10 samples for edges, and we later report sensitivity to different discretization resolutions in Sec.~\ref{sec:results}.}

\subsubsection{Attributes Information}
To enable our model to learn better BRep representations, we extract face and edge attributes from BRep models, as AAGNet does~\cite{wu2024aagnet}.
Specifically, for each face, we extract a 10-dimensional attribute vector $f_{\text{attr}} \in \mathbb{R}^{10}$, including surface types (plane, cylinder, sphere, etc.), surface area, and the 3D centroid coordinates.
The surface types have 6 types and are encoded using a one-hot scheme.
We also compute a 6-dimensional axis-aligned bounding box (AABB) vector $f_{\text{aabb}} \in \mathbb{R}^6$ to describe the spatial bounds.
Similarly, for each edge, we extract a 14-dimensional attribute vector $e_{\text{attr}} \in \mathbb{R}^{14}$, including curve type (line, circle, ellipse, etc.), length, and convexity (concave, convex, or smooth).
\blu{Finally, face/edge attributes as well as AABB parameters are standardized using dataset-level mean/std statistics.}
Tab.~\ref{tab:attr-info} shows the details of the attributes of faces and edges.
%

%

\begin{figure*}[t]
 \centering
		\begin{overpic}[width=0.99\linewidth]{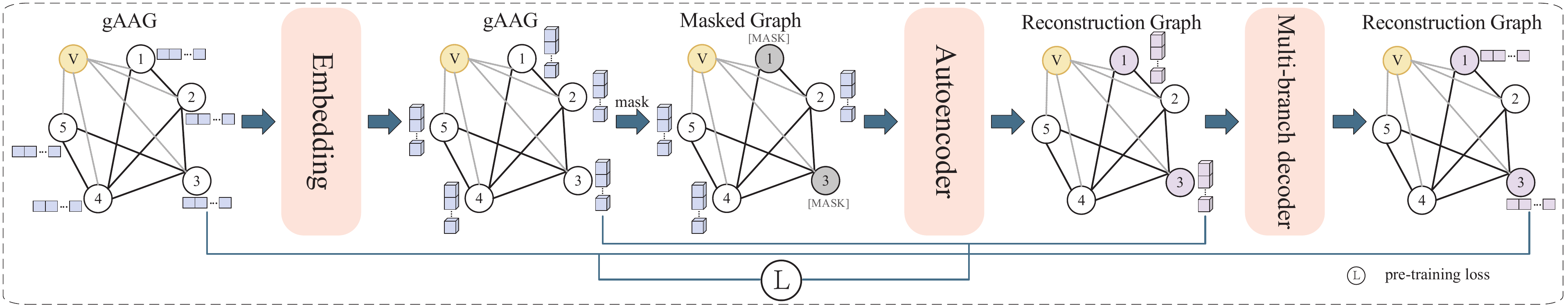}
			{
		    }
		\end{overpic}
		 \caption{
         The workflow of our pre-training stage.
         After we convert the input BRep model into a gAAG, our model embeds the geometric information of faces and edges into a 256-dimensional space.
         Then we randomly mask some nodes of the gAAG and reconstruct their information using an autoencoder.
         A multi-branch decoder is adopted to recover the geometric information of BRep models.
         The reconstructed 256-dimensional features and geometric information are used in the loss to pre-train our model.
			 }
		 \label{fig:pre-training}
\end{figure*}

\begin{figure*}[t]
 \centering
		\begin{overpic}[width=0.99\linewidth]{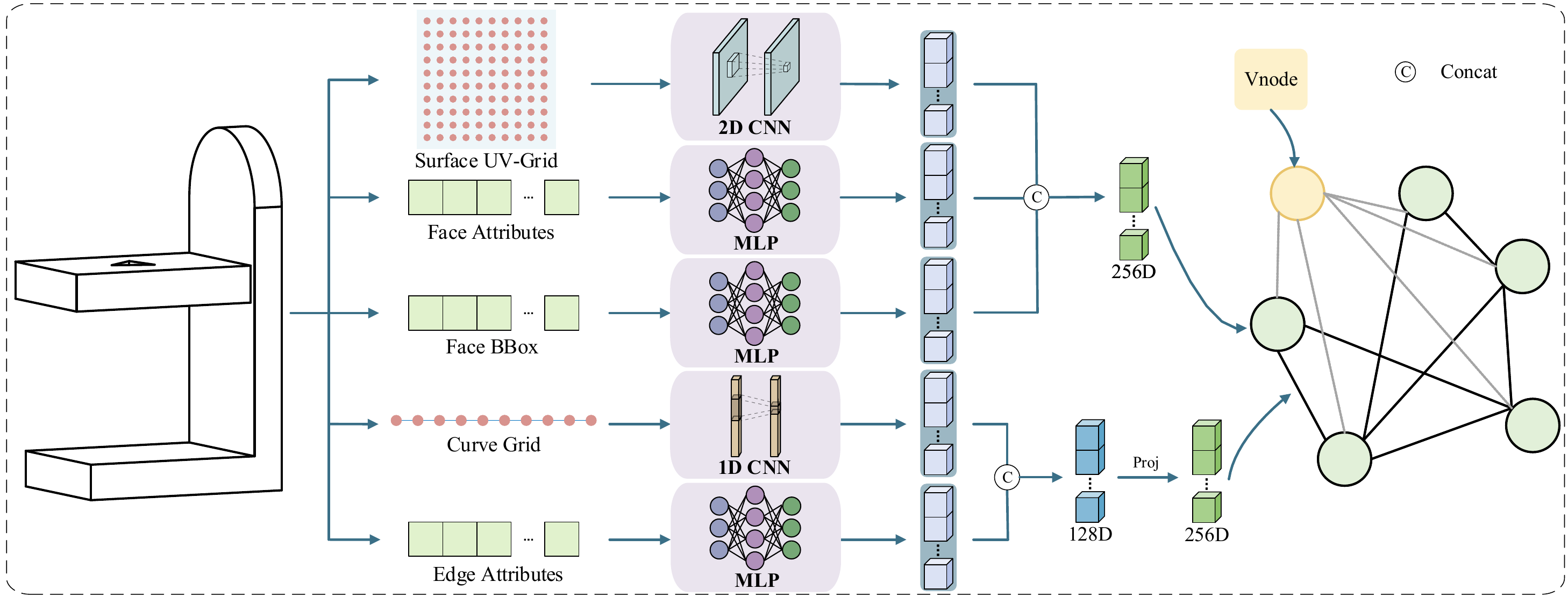}
			{
		    }
		\end{overpic}
		 \caption{
        Our BRep embedding. 
       Various networks (such as 2D CNNs, MLPs, and 1D CNNs) are used to embed face and edge information.
        The face feature vectors are concatenated to form a 256-dimensional vector, which serves as the graph node feature.
        Similarly, the edge feature vectors are concatenated and projected to produce a 256-dimensional vector that serves as the graph edge feature.
			 }
		 \label{fig:BRep-embedding}
\end{figure*}

\begin{table}[!t]
\scriptsize
\setlength{\lightrulewidth}{0.3pt}
\begin{tabularx}{0.48\textwidth}{p{0.7cm}p{1.1cm}p{3.8cm}p{2cm}}
    \hline
    \textbf{Entity} & \textbf{Attribute} & \textbf{Description} & \textbf{Dimension} \\
    \midrule
    \multirow{2}{*}{Face} & \multirow{2}{*}{Type} & plane, cylinder, cone, & \multirow{2}{*}{6 (one-hot)}\\
     & &  sphere, torus, or NURBS & \\
    \midrule
    Face & Area & The area of the face & 1 \\
    \midrule
    Face & Centroid &  The centroid coordinates of the face & 3 \\
    \midrule
    \multirow{2}{*}{Face} & \multirow{2}{*}{BBOX} &  The centroid coordinates of the BBOX, & \multirow{2}{*}{3+3} \\
     & &  the half size of the BBOX & \\
    \hline
    \multirow{2}{*}{Edge} & \multirow{2}{*}{Type} & circle, line, ellipse, parabola, hyperbola, & \multirow{2}{*}{10 (one-hot)}\\
     & &  closed, Bezier, B-Spline, RBS, or offset & \\
    \midrule
    Edge & Length & The length of the edge & 1 \\
    \midrule
    Edge & Convexity & concave, convex, or smooth & 3 (one-hot) \\
    \hline
\end{tabularx}
\caption{\blu{Face/edge attribute groups used in gAAG construction with their dimensionality (one-hot or scalar), which together form the node/edge input features.}
}
\label{tab:attr-info}
\centering
\end{table}

\subsection{Pre-training}\label{sec:pre-training}
To improve our model's generalization and remove its reliance on labeled data, we pre-train it with MAE, a self-supervised learning method (Fig.~\ref{fig:pre-training}).
%
%
%
Given a gAAG representing a BRep model, our method first embeds node and edge features into high-dimensional spaces (Sec.~\ref{sec:embedding}).
Then, inspired by the successful development of GraphMAE~\cite{hou2022graphmae}, we randomly mask certain embedded features of the nodes for the encoder and reconstruct the masked features by the decoder without changing the graph topology to train our masked autoencoder (Sec.~\ref{sec:masked-autoencoder}) with a tailored loss function (Sec.~\ref{sec:loss-4-per-training}).
%

\subsubsection{BRep Embedding}\label{sec:embedding}
Our BRep feature encoder consists of a 2D CNN for face geometry, a 1D CNN for edge geometry, and three MLPs for face bounding box, face attributes, and edge attributes (Fig.~\ref{fig:BRep-embedding}).
%
\blu{The details of these encoders are shown in the supplementary materials.} 
%
To better leverage global information, we introduce a virtual node to gAAG that is connected to all nodes, whose node embedding is a 256-dimensional vector, and the edge embeddings are 128-dimensional vectors.
Therefore, the gAAG is defined as $\mathcal{G} = (\mathcal{V}, \mathcal{E}, \mathbf{X}, \mathbf{X_E})$, where $\mathcal{V}$ is the set of nodes containing the virtual node, $\mathcal{E}$ is the set of edges containing the edges between virtual node and other nodes, $\mathbf{X} \in \mathbb{R}^{|\mathcal{V}| \times 256}$ is the face embeddings and $\mathbf{X_E} \in \mathbb{R}^{|\mathcal{E}| \times 256}$ is the edge embeddings.



\begin{figure*}[t]
 \centering
		\begin{overpic}[width=0.99\linewidth]{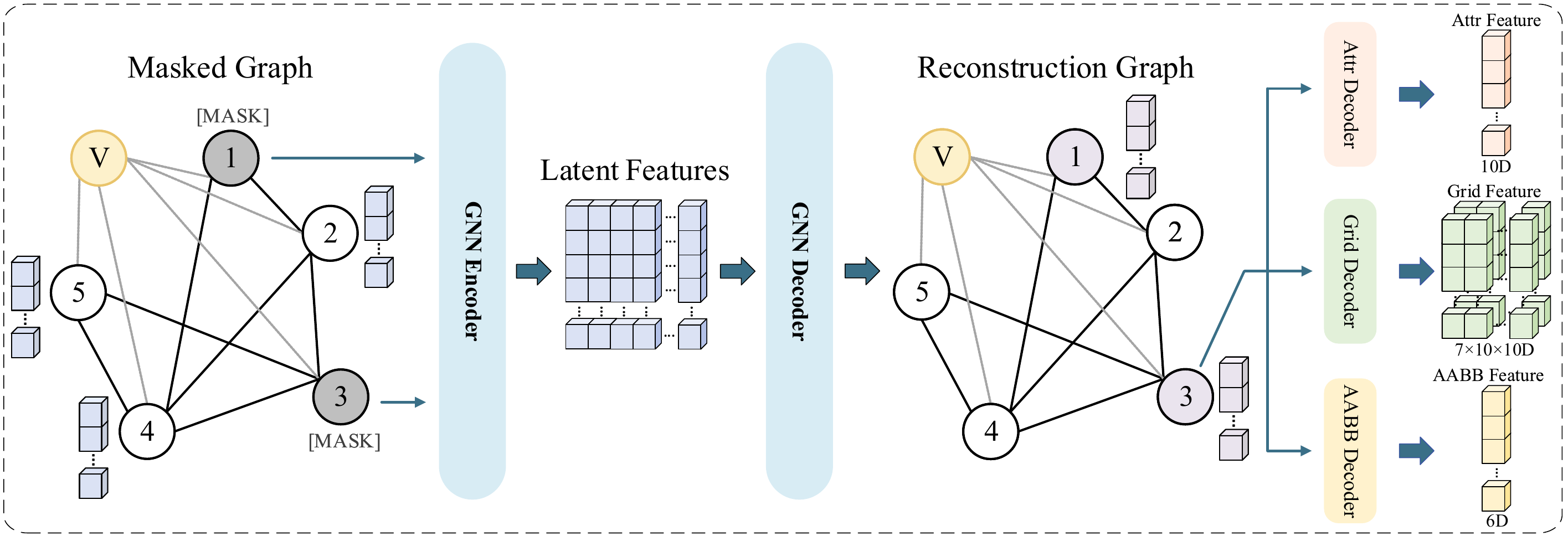}
			{
		    }
		\end{overpic}
		 \caption{
         Given a graph with some nodes' features masked, our model first applies a GNN encoder (MPNNs) to extract the graph's latent representation, and then a GNN decoder (MPNNs) to reconstruct the masked node features.
         Later, a multi-branch decoder is implemented to decode the geometric and attribute information of the BRep face from the reconstructed node feature.
			 }
		 \label{fig:mased-autoencoder}
\end{figure*}

\subsubsection{Masked Autoencoder}\label{sec:masked-autoencoder}
%
Since graph structures enable our model to leverage the structural dependencies of BRep to learn CAD representations more effectively, we implement the encoder and decoder of our masked autoencoder using the Message Passing Neural Network (MPNN)~\cite{wu2024aagnet} (\blu{see the supplementary materials for more details}).
%
%
After iteratively performing a certain number of message passings, each node will eventually contain information from the entire graph.
As with AAGNet, to capture richer neighborhood information, our model does not use a single aggregation function; instead, it introduces a PNAlayer~\cite{corso2020principal} to fuse multiple aggregation results.
Specifically, given a gAAG $\mathcal{G}$ with some masked node features as input, our encoder first maps $\mathcal{G}$ into a latent space without changing the features' dimensionality and the topology of $\mathcal{G}$.
Then, the decoder maps the latent representation back to reconstruct the input graph information, allowing our model to learn a representation of the BRep model (see Fig.~\ref{fig:mased-autoencoder}).
%

%

\subsubsection{Loss for pre-training}\label{sec:loss-4-per-training}
In the pre-training stage, our goal is to enable our model to learn BRep representations by reconstructing BRep models from masked node features in $\mathcal{G}$.
The loss consists of the following four aspects: (1) the masked node feature $\mathbf{X}$ in $\mathcal{G}$, (2) face geometric representation $f_{\text{geom}}$, (3) face attributes information $f_{\text{attr}}$, and (4) face bounding box $f_{\text{aabb}}$:
\begin{equation}\label{equ:total-loss}
    \mathcal{L} = \alpha\mathcal{L}_\text{feat} + \beta\mathcal{L}_\text{geom} + \gamma\mathcal{L}_\text{attr} + \delta\mathcal{L}_\text{aabb},
\end{equation}
where $\mathcal{L}_\text{feat}$ is used to reconstruct the masked node feature in $\mathcal{G}$, $\mathcal{L}_\text{geom}$, $\mathcal{L}_\text{attr}$, and $\mathcal{L}_\text{aabb}$ are applied to reconstruct $f_{\text{geom}}$, $f_{\text{attr}}$, and $f_{\text{aabb}}$, respectively.
\blu{$\mathcal{L}_\text{feat}$, $\mathcal{L}_\text{attr}$, and $\mathcal{L}_\text{aabb}$ are defined as Mean Squared Error (MSE) (see details in the supplementary material).}
In our experiments, we set $\alpha = 0.4$, $\beta = 0.36$, $\gamma = 0.12$, and $\delta = 0.12$.

\begin{figure}[t]
 \centering
		\begin{overpic}[width=0.99\linewidth]{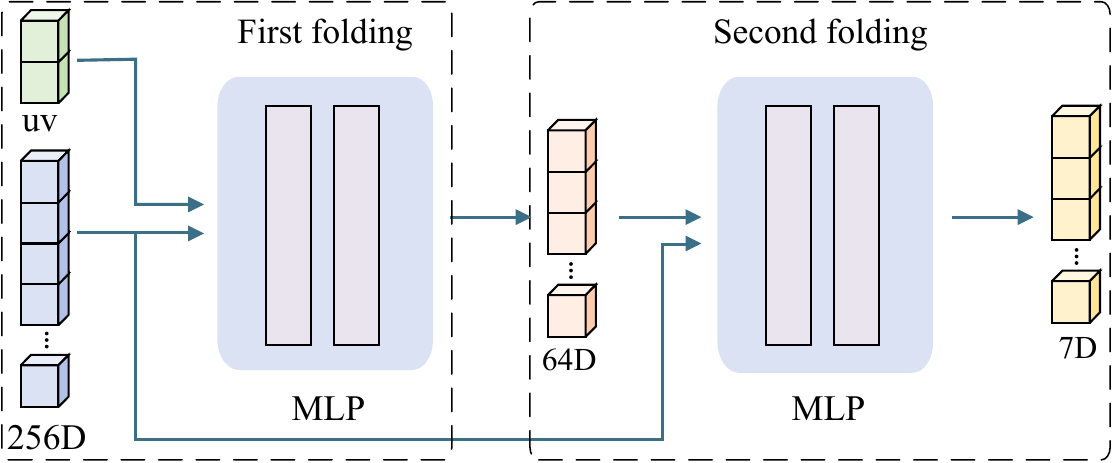}
			{
		    }
		\end{overpic}
		 \caption{
         Our geometric information decoder with two folding stages.
         %
         %
			 }
		 \label{fig:foldingnet}
\end{figure}

\paragraph{Geometric information term}
\blu{Instead of using a simple MLP decoder for surface reconstruction, we employ a FoldingNet~\cite{yang2018foldingnet} decoder with two folding stages to predict the input BRep geometry (see more discussions in Sec.~\ref{sec:ablation}).} 
%
Specifically, given a node feature $\mathbf{x}_i\in \mathbb{R}^{256}$ and a point on the UV parameter domain $\mathbf{t}\in \mathbb{R}^{2}$, our decoder concatenates these two vectors as the input of the first folding stage and outputs a 64-dimensional vector (Fig.~\ref{fig:foldingnet}).
In the second folding stage, $\mathbf{x}_i$ and the 64-dimensional vector after the first folding stage are concatenated as the input, and a 7-dimensional vector $\hat{f}_{\text{geom}, i}=(\hat{X}_i^s, \hat{Y}_i^s, \hat{Z}_i^s, \hat{N}^s_{x, i}, \hat{N}^s_{y, i}, \hat{N}^s_{z, i}, \hat{T}_i^s)$ is the predicted geometric information of our decoder.
%
\blu{The details of this decoder are shown in the supplementary materials.}

Since the geometric information consists of 3D coordinates, normals, and whether the point is in the trimmed area or not, the error definitions between the reconstruction $\hat{f}_\text{geom}$ and the ground truth $f_\text{geom}$ of these three parts are different:
%
\begin{equation}\label{equ:geom-loss}
    \mathcal{L}_\text{geom} = \beta_1 \mathcal{L}_\text{coord} + \beta_2 \mathcal{L}_\text{norm} + \beta_3\mathcal{L}_\text{trim}, 
\end{equation}
where
\begin{equation}\label{equ:coord-loss}
    \mathcal{L}_\text{coord} =  \frac{1}{n}\sum_{i=0}^{n-1}\| \mathbf{p}_i - \hat{\mathbf{p}}_i \|_2^2,
\end{equation}
\begin{equation}\label{equ:norm-loss}
    \mathcal{L}_\text{norm} =  \frac{1}{n}\sum_{i=0}^{n-1}(1 -  \frac{\mathbf{n}_i \cdot \hat{\mathbf{n}}_i }{ \| \mathbf{n}_i\| \cdot  \|\hat{\mathbf{n}}_i \|}) ,
\end{equation}
\begin{equation}\label{equ:trim-loss}
    \mathcal{L}_\text{trim} =  \frac{1}{n}\sum_{i=0}^{n-1}(-T_i^s\log(\hat{T_i^s}) - (1-T_i^s)\log(1-\hat{T_i^s}) ).
\end{equation}
The vector $ \hat{\mathbf{p}}_i = (\hat{X}_i^s, \hat{Y}_i^s, \hat{Z}_i^s)$ is the predicted 3D coordinate, $\hat{\mathbf{n}}_i = (\hat{N}^s_{x, i}, \hat{N}^s_{y, i}, \hat{N}^s_{z, i})$ is the predicted surface normal, $\hat{T_i^s}$ is the predicted trimmed state, $n$ is the number of sampling points, and $\mathbf{p}_i$, $\mathbf{n}_i$, and $T_i^s$ are the ground truth.
In our experiments, we set $\beta_1 = 0.4$, $\beta_2 = 0.3$ and $\beta_3 = 0.3$.

\subsection{Fine-tuning}\label{sec:fine-tuning}
To adapt our general B-Rep representations learned during pre-training for specific downstream tasks, we employ a fine-tuning stage.
%
%
In this stage, various amounts of labeled data are used to retrain our model.
The BRep embedding module and the encoder of our masked autoencoder are retained, and their weights are used for initialization.
The rest of our model is discarded entirely.
A task-specific classification head is appended after these modules to enable our network to be used for machining feature recognition.
%
%


\subsubsection{Task-specific classification head}
We append a task-specific classification head, which is a three-layer Multi-Layer Perceptron (MLP), to the MPNN encoder. 
\blu{The detailed architecture is shown in the supplementary materials.} 
%
%
%
%
This MLP maps the node embeddings (256-dimensional) to a logit vector of size $n_c$, where $n_c$ is the total number of predefined machining feature classes (e.g., hole, slot, pocket, fillet, etc.).
When reasoning, we use the softmax function to convert the logit vector into a probability distribution, and then determine the type of the corresponding BRep face based on the size of each component.
In our experiments, we set $n_c=25$.

\subsubsection{Loss for fine-tuning}
The entire model, comprising the MPNN encoder and the newly added classification head, is trained on a limited amount of data with face-level labels. 
We adopt the standard cross-entropy loss function to measure the dissimilarity between the predicted class probabilities and the ground-truth one-hot labels for each face in a given CAD model: 
\begin{equation}\label{equ:fine-tuning-loss}
    \mathcal{L}_\text{fine} =  -\sum_{i=0}^{n_f-1}\sum_{j=0}^{n_c-1}y_{i, j}\log(\text{Softmax}(z_{i, j})),
\end{equation}
where $n_f$ is the number of faces in a given CAD model, $z_{i, j}$ is the $j\text{th}$ component of the output logit vector from the $i\text{th}$ face in the input CAD model.
$y_{i, j}$ is the ground truth (1 if face $i$ belongs to $j\text{th}$ surface type, and 0 otherwise).
%


\section{Results and Comparisons}\label{sec:results}
Our model is implemented in Python, and all experiments were conducted on a high-performance computing node equipped with a 16-core 8358P CPU, 64 GB of system memory, and a single NVIDIA A800 GPU. 

\subsection{Dataset}
\paragraph{Dataset construction}
Our pre-training dataset includes ABC~\cite{koch2019abc}, MFInstSeg~\cite{wu2024aagnet}, Fusion 360 Gallery~\cite{lambourne2021brepnet}, CADSynth~\cite{zhang2024brepmfr}, MFCAD2~\cite{colligan2022hierarchical}, and MFCAD~\cite{cao2020graph}. 
For ABC, we select a subset of models and then apply two filters: retain single-entity parts and restrict the number of faces to at most 1,000 to ensure stable B-Rep processing. 
After filtering, we obtain 44,773 valid entities with diverse surface types and topological structures. 
For the other datasets, we similarly remove models with invalid or non-closed topology, missing edge-loop information, or failures during parameterization/conversion (\blu{see more detailed failure examples in the supplementary materials.}). 
%
Finally, we extract 62,494 processing feature models from the MFInstSeg dataset; 35,252 real-world CAD models from the Fusion 360 Gallery; 95,396 and 59,450 models with labeled face-level annotations from the CADSynth and MFCAD2 datasets, respectively; and 15,488 models with semantic information for 24 machining feature classes from the MFCAD dataset.
%

\paragraph{Evaluation settings}
We evaluate BRepMAE under three complementary settings.
\begin{enumerate}
    \item To make a fair comparison with full-supervised baselines, we pre-train and fine-tune on the \emph{same} dataset when reporting the label-ratio experiments.
Specifically, the MFInstSeg~\cite{wu2024aagnet}, CADSynth~\cite{zhang2024brepmfr}, and MFCAD2~\cite{colligan2022hierarchical} datasets are first divided into the standard 80/10/10 splits.
During the self-supervised stage, we use only the 80\% training split as unlabeled data, and we merge the official validation and test splits (10\%+10\%) to monitor the reconstruction loss because no downstream evaluation is conducted at this time.
After pre-training, the original 80/10/10 split is restored and used for fine-tuning (train/validation/test), ensuring that the evaluation protocol remains unchanged.
The fully supervised learning method uses the same proportion of training data for training as the comparison result.
\item To compare with BRep-BERT~\cite{lou2023brep}, we strictly follow its protocol by pre-training on the Fusion 360 Gallery~\cite{lambourne2021brepnet} dataset and evaluating 10-way few-shot segmentation on MFCAD~\cite{cao2020graph}.
\item We retain our large-scale corpus of over 312,000 CAD models aggregated from ABC~\cite{koch2019abc}, MFInstSeg, Fusion 360 Gallery, CADSynth, MFCAD2, and MFCAD.
For ABC, we retain 44,773 valid single-entity parts (each with at most 1,000 faces), and we apply the same validity check (removing open surfaces, missing loops, and failed parameterization) to the other datasets.
This corpus is split into 80\% training and 20\% validation and is used for pre-training before performing 24-way few-shot evaluation on MFInstSeg, CADSynth, and MFCAD2.
\end{enumerate}

\subsection{Pre-training}
\paragraph{Training settings}
In the pre-training stage, we use the AdamW optimizer~\cite{loshchilov2017decoupled} with a cosine learning rate scheduler for stable training.
The initial learning rate is set to $10^{-4}$, the batch size is $256$, the max epochs are $150$, and the masking ratio is $80\%$.
\blu{Under the above setting, the pre-training takes about 11 hours on a single NVIDIA A800 GPU.}

\paragraph{Efficiency profile}
\blu{To better quantify the computational cost of BRepMAE, we report the model size and efficiency in Tab.~\ref{tab:efficiency}.}
\blu{We distinguish the full pre-training model (encoder + decoders) from the encoder-only backbone that is typically used for downstream tasks and deployment.}
\blu{We report Params, FLOPs per graph (averaged on graphs with $N\approx30$ nodes), and inference latency. FLOPs are estimated by combining torch-level FLOPs counting with an additional estimate of DGL graph aggregation operations, and latency is measured on the same single NVIDIA A800 GPU.}

\begin{table}[t!]
\centering
\scriptsize
\setlength{\tabcolsep}{2pt}
\begin{tabular}{@{}>{\raggedright\arraybackslash}p{0.28\linewidth}ccc@{}}
\toprule
\textbf{Model} & \textbf{Params (M)} & \textbf{FLOPs (G)} & \textbf{Latency (ms)} \\
\midrule
Full pre-train & 8.03 & 7.26 & 1.17 \\
Encoder only & 3.66 & 2.62 & 0.73 \\
\bottomrule
\end{tabular}
\caption{\blu{Efficiency profile of BRepMAE.}} 
\label{tab:efficiency}
\end{table}

\begin{figure*}[t]
 \centering
		\begin{overpic}[width=0.93\linewidth]{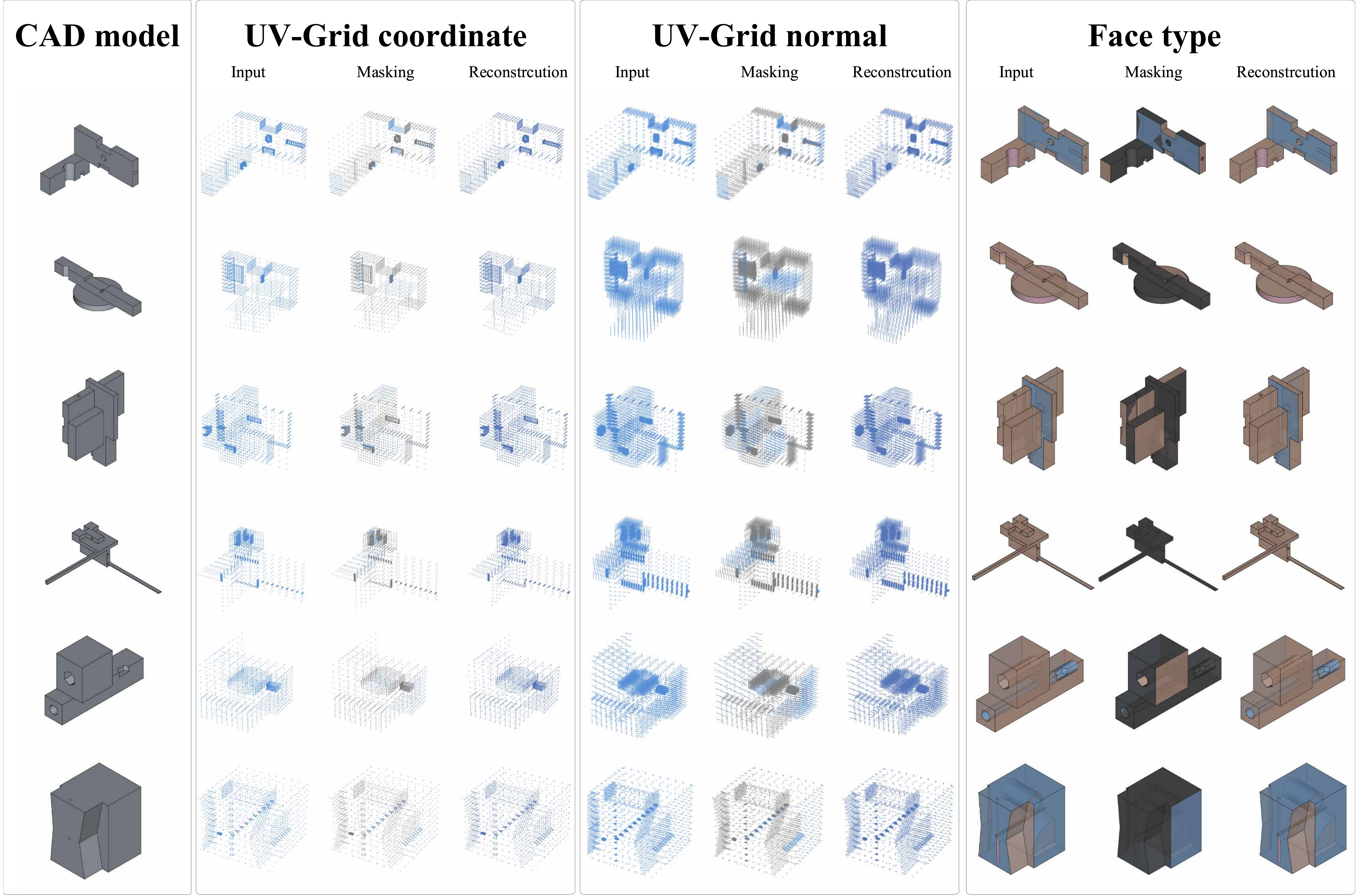}
			{
		    }
		\end{overpic}
		 \caption{
         Representative reconstruction results.
         We set the masked parts of the CAD models to black.
         In the face type column, we paint different face types with different colors.
			 }
		 \label{fig:reconstruction}
\end{figure*}
\begin{figure*}[!t]
 \centering
 \includegraphics[width=0.93\linewidth]{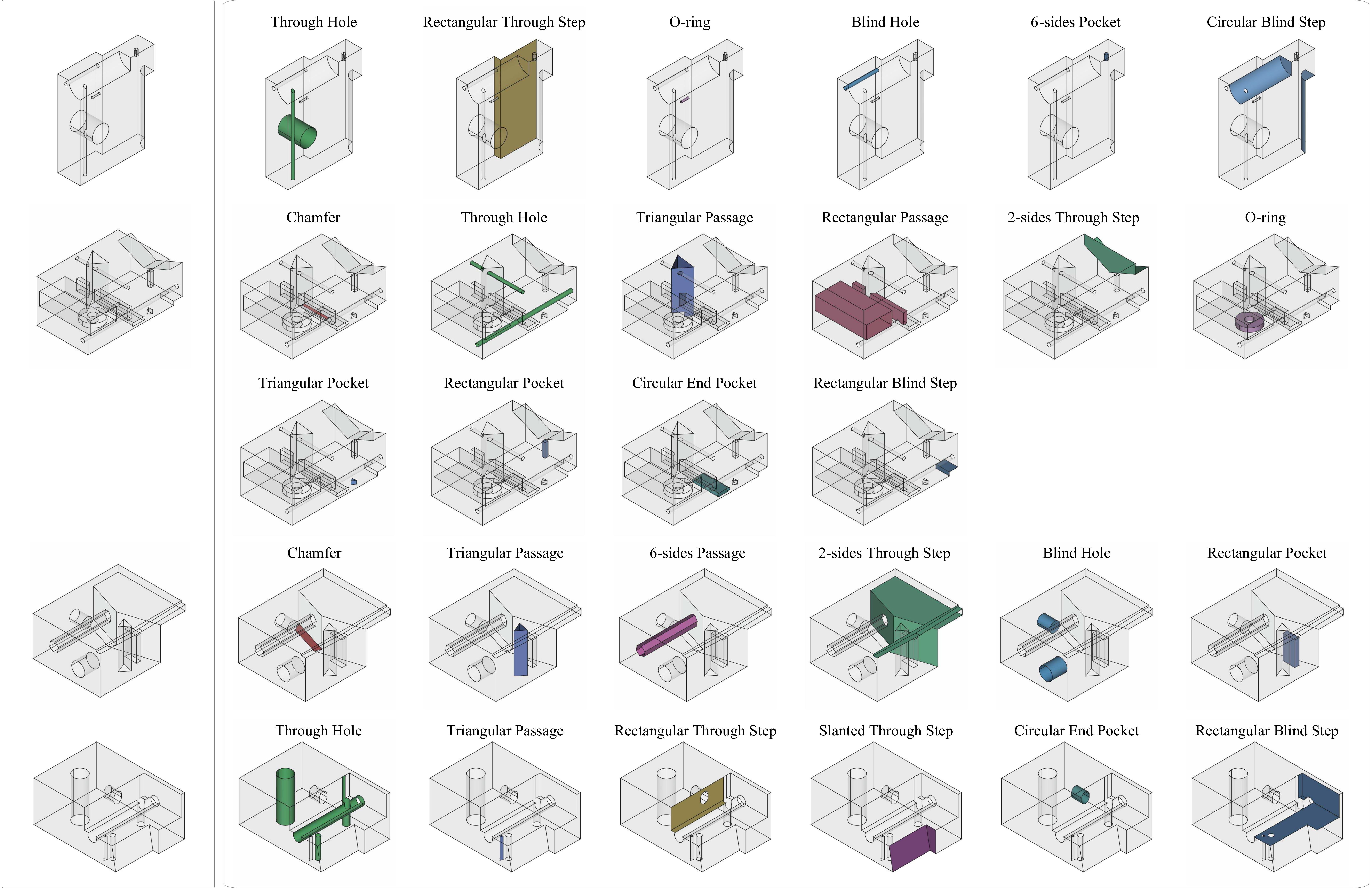}
 \caption{\blu{Representative machining feature recognition visualizations. Different machining features are shown in different colors.}}
 \label{fig:machining-feature}
\end{figure*}

\paragraph{Reconstruction}
\blu{We present representative reconstruction results (UV-grid coordinates, normals, and face types) generated by our pre-training model in Fig.~\ref{fig:reconstruction}, with additional examples provided in the supplementary material.}
%
Our model accurately reconstructs masked CAD surfaces and infers their normals from unmasked faces. More importantly, it correctly reasons about masked face types, demonstrating that it learns high-level BRep representations, confirming the effectiveness of pre-training on BRep. 

\subsection{Fine-tuning}
For the fine-tuning stage, we retain the pre-trained encoder of the autoencoder and replace the decoder with a task-specific classification head implemented as a three-layer MLP with dimensions $[d_\text{latent}, 1024, 256, N_\text{clas}]$, where $d_\text{latent}=256$ is the encoder output dimension and $N_\text{clas}$ is the number of machining feature classes.
We employ the AdamW optimizer with an initial learning rate of $5\times 10^{-4}$, weight decay of $10^{-4}$.
The training is conducted for up to $50$ epochs, with early stopping (patience of $5$ epochs) to prevent overfitting.
We also perform an ablation study by freezing the encoder of our autoencoder to demonstrate the necessity of the fine-tuning stage.
The results are shown in Tab.~\ref{tab:ablation_acc}.
\blu{We further show representative MFR results in Fig.~\ref{fig:machining-feature}. More examples are provided in the supplementary material.}
%

%
%
%
%

\begin{figure}[t]
 \centering
 \begin{minipage}{0.485\linewidth}
  \centering
  \includegraphics[width=\linewidth]{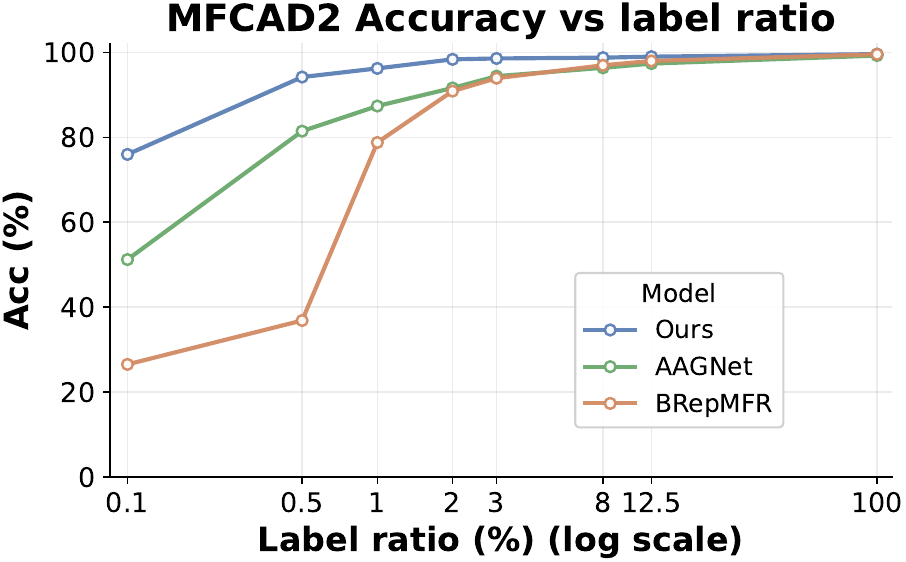}
 \end{minipage}%
 \hfill%
 \begin{minipage}{0.485\linewidth}
  \centering
  \includegraphics[width=\linewidth]{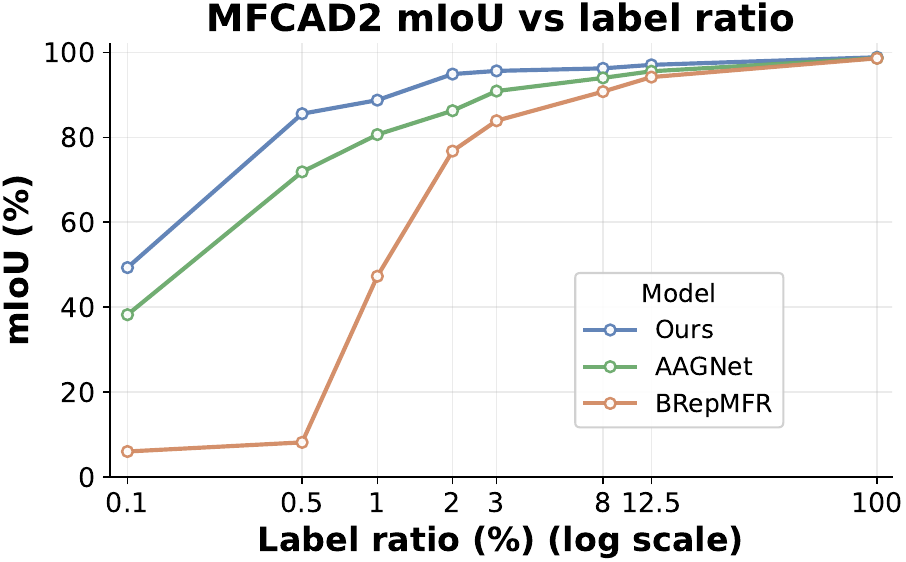}
 \end{minipage}
 \par\vspace{0.25em}

 \begin{minipage}{0.485\linewidth}
  \centering
  \includegraphics[width=\linewidth]{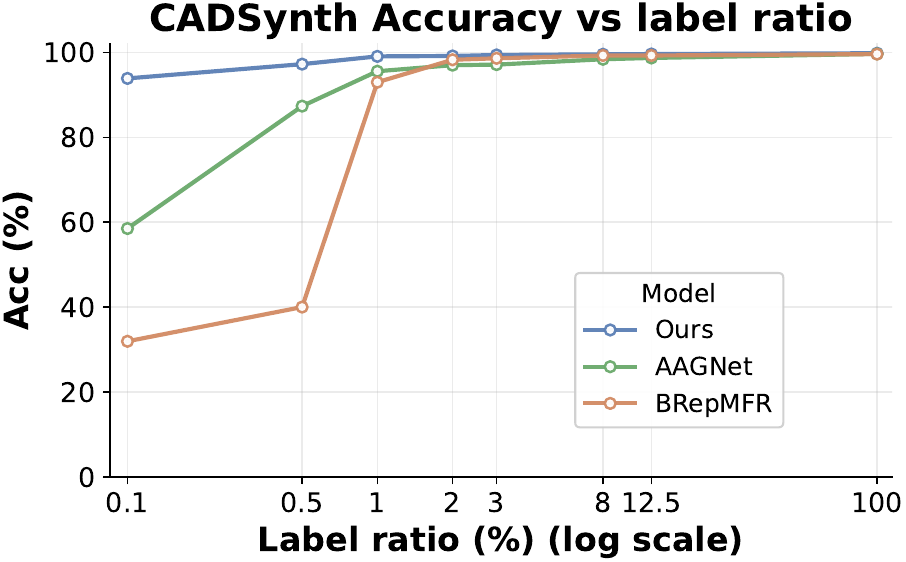}
 \end{minipage}%
 \hfill%
 \begin{minipage}{0.485\linewidth}
  \centering
  \includegraphics[width=\linewidth]{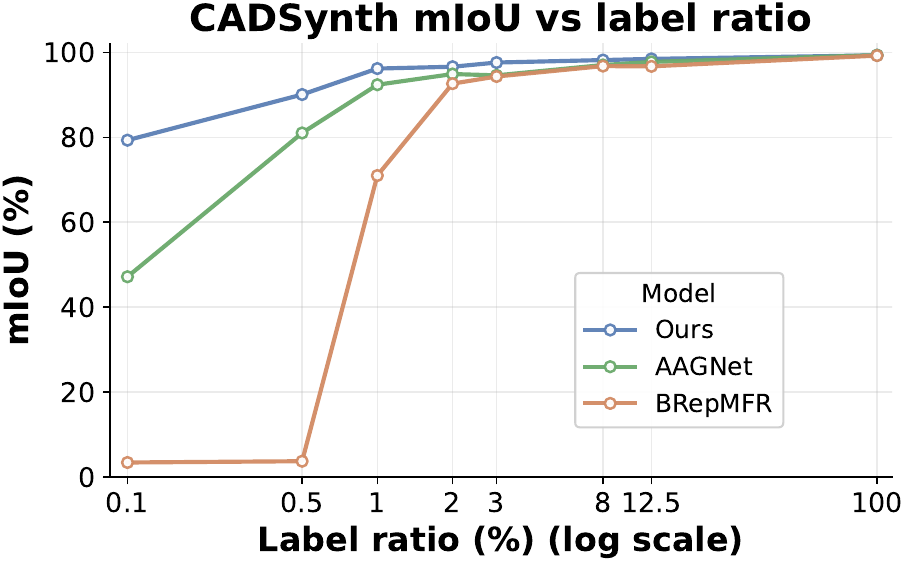}
 \end{minipage}
 \par\vspace{0.25em}

 \begin{minipage}{0.485\linewidth}
  \centering
  \includegraphics[width=\linewidth]{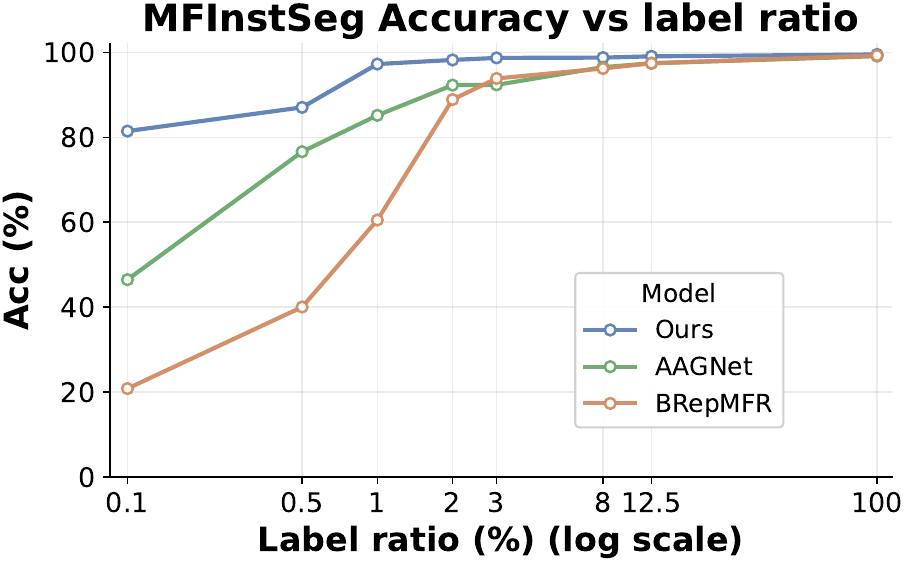}
 \end{minipage}%
 \hfill%
 \begin{minipage}{0.485\linewidth}
  \centering
  \includegraphics[width=\linewidth]{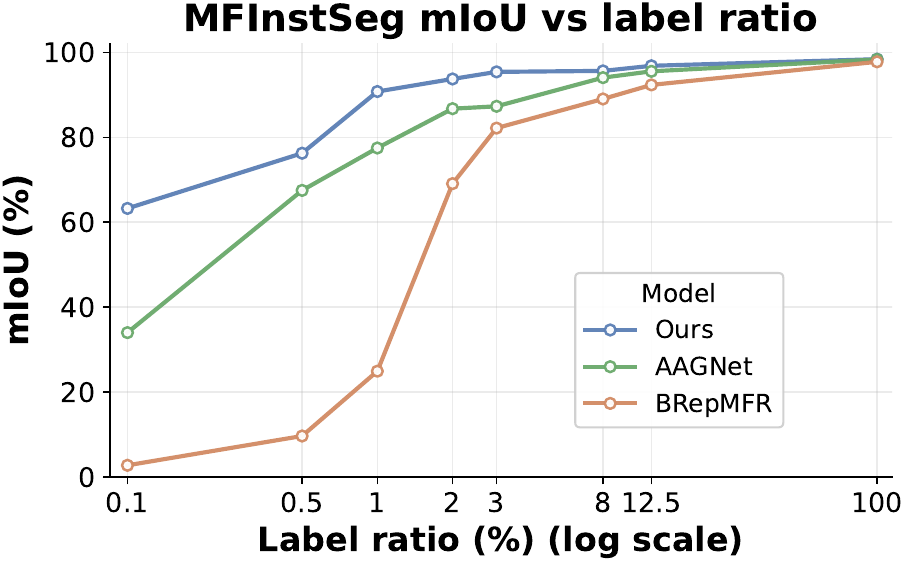}
 \end{minipage}
 \caption{\blu{Trend visualizations across three datasets (x-axis is log-scaled label ratio). The top/middle/bottom rows correspond to MFCAD2/CADSynth/MFInstSeg, respectively.}}
 \label{fig:trend_all}
\end{figure}

\subsection{Comparisons with full-supervised methods}
%
The comparison with AAGNet~\cite{wu2024aagnet} and BRepMFR~\cite{zhang2024brepmfr} is performed on the MFCAD2, CADSynth, and MFInstSeg datasets.
%
They are trained using their original implementations and default hyperparameters for fair comparisons. 
We focus on the face semantic segmentation task, which is central to MFR. 
The performance is evaluated using face-level accuracy (Acc, \%) and mean Intersection over Union (mIoU) (\blu{see definitions in the supplementary material}). 
%
%
In practice, we first pre-train our model on this dataset, treating it as unlabeled data.
Then, to evaluate our model's performance with limited labeled data, we use different proportions (ranging from 0.1\% to 100\%) of labeled data in the dataset to fine-tune (or train) our model (or a full-supervised method) for comparison.
%

\blu{The comparison results are shown in Fig.~\ref{fig:trend_all}.
The lower the proportion of labeled data, the more pronounced the advantage of our method becomes.}
These results indicate that during the pre-training stage, our model extracts effective latent features from the unlabeled portion of the dataset, enabling better performance than full-supervised learning methods under limited labeled-data conditions.
Specifically, when the distribution of unlabeled and labeled data is equal, our method can effectively leverage labeled information for downstream tasks.

\subsection{Few-shot Learning}

\subsubsection{Comparison with BRep-BERT}
We first compare with BRep-BERT~\cite{lou2023brep} under its official protocol: the model is pre-trained on Fusion 360 Gallery and then evaluated on the MFCAD dataset using a fixed 10-way setup with either 10-shot or 20-shot supports. 
Tab.~\ref{tab:brepbert} reproduces the results reported in~\cite{lou2023brep}, which cover point-based baselines (Point-BERT~\cite{yu2022point}, Point-MAE~\cite{pang2023masked}, Point-M2AE~\cite{zhang2022point}), the UV-Net~\cite{jayaraman2021uv} surface model, the BRepNet~\cite{lambourne2021brepnet} graph model, and two BRep-BERT variants (a standard Transformer backbone denoted as BRep-BERT(T) and the final BRep-BERT). 
%
Under the same conditions, our BRepMAE outperforms the previous methods. 
%
%
\blu{To control for potential differences in input representations, we additionally evaluate an input-aligned variant, termed Ours (BRep-BERT-aligned gAAG), where we match BRep-BERT's graph structure and input feature configuration (both nodes and edges).}
\blu{Under this input-aligned setting, our method still outperforms BRep-BERT (Tab.~\ref{tab:brepbert}), indicating that the improvement does not solely come from richer inputs.}
\blu{Meanwhile, the full gAAG setting achieves the best results, suggesting that richer geometric/attribute signals can further benefit performance.}
This shows that our MAE-based BRep learning framework can extract more effective features for downstream tasks.

\begin{table}[t]
\centering
\scriptsize
\setlength{\tabcolsep}{4pt}
\begin{tabular}{p{0.44\linewidth}cccc}
\toprule
\multirow{2}{*}{\textbf{Model}} & \multicolumn{2}{c}{\textbf{10-shot}} & \multicolumn{2}{c}{\textbf{20-shot}} \\
\cmidrule(lr){2-3} \cmidrule(lr){4-5}
& \textbf{Acc.} & \textbf{IoU} & \textbf{Acc.} & \textbf{IoU} \\
\midrule
Point-BERT & 31.06 & 9.40 & 39.04 & 10.37 \\
Point-MAE & 32.36 & 9.61 & 40.04 & 11.06 \\
Point-M2AE & 34.86 & 10.70 & 42.34 & 12.57 \\
UV-Net & 27.21 & 8.19 & 30.04 & 8.31 \\
BRepNet & 35.03 & 10.19 & 39.83 & 11.53 \\
BRep-BERT (T) & 50.05 & 16.05 & 53.04 & 18.10 \\
BRep-BERT & 54.05 & 18.82 & 58.13 & 21.05 \\
\midrule
\textbf{Ours \blu{(full gAAG)}} & \textbf{56.85} & \textbf{42.80} & \textbf{81.58} & \textbf{73.54} \\
\blu{Ours (BRep-BERT-aligned gAAG)} & \blu{55.47} & \blu{31.51} & \blu{80.86} & \blu{61.76} \\
\bottomrule
\end{tabular}
\caption{Few-shot machining feature segmentation on MFCAD following the BRep-BERT protocol (10-way classification with 10-shot and 20-shot supports). The first seven rows are the results reported by BRep-BERT in percentages.
\blu{Ours (BRep-BERT-aligned gAAG) matches BRep-BERT's graph construction and node/edge feature specification, while keeping the same training protocol.}}
\label{tab:brepbert}
\end{table}

\subsubsection{24-way few-shot }
We also exploit our 312K mixed-source dataset to pre-train BRepMAE and then evaluate it with few-shot learning tasks on different sub-datasets. 
Following a 24-way setting, each dataset (MFInstSeg, CADSynth, and MFCAD2) is split into train/val/test with an 80/10/10 ratio. 
We construct few-shot episodes by sampling support/query sets from the test split with 10-shot and 20-shot settings, while the validation split is used for model selection.
Tab.~\ref{tab:24way} reports the obtained accuracies and IoUs. 
Despite the larger number of classes and cross-domain shift, BRepMAE still delivers promising performance. 
These results show that the proposed self-supervised framework can be used on large, heterogeneous datasets and yields general features for downstream tasks.

\begin{table}[t]
\centering
\scriptsize
\setlength{\tabcolsep}{12pt}
\begin{tabular}{lcccc}
\toprule
\multirow{2}{*}{\textbf{Dataset}} & \multicolumn{2}{c}{\textbf{10-shot}} & \multicolumn{2}{c}{\textbf{20-shot}} \\
\cmidrule(lr){2-3} \cmidrule(lr){4-5}
& \textbf{Acc.} & \textbf{IoU} & \textbf{Acc.} & \textbf{IoU} \\
\midrule
MFInstSeg & 59.18 & 43.97 & 77.85 & 63.75 \\
CADSynth & 63.70 & 54.62 & 82.93 & 77.48 \\
MFCAD2 & 55.19 & 42.09 & 75.63 & 62.02 \\
\bottomrule
\end{tabular}
\caption{24-way few-shot performance (\%) after pre-training on the 312K mixed-source corpus. Each dataset is evaluated with 10-shot and 20-shot tasks sampled using an 80/10/10 split.}
\label{tab:24way}
\end{table}

\begin{table}[t]
\centering
\setlength{\tabcolsep}{1.5pt}
\begin{tabular}{lccccccccc}
\toprule
\textbf{Configuration} & \textbf{0.1\%} & \textbf{0.5\%} & \textbf{1\%}  & \textbf{3\%} & \textbf{8\%} & \textbf{12.5\%} & \textbf{100\%} \\
\midrule
w/o VN & 82.34 & 91.85 & 96.42  & 98.15 & 98.46 & 98.76 & 99.55 \\
Folding(1) & 83.10 & 90.78 & 94.02 & 97.43 & 98.45 & 98.67 & 99.43 \\
Folding(3) & 84.97 & 94.69 & 96.67 & 97.95 & 98.66 & 99.05 & 99.50 \\
\blu{FoldingNet $\rightarrow$ MLP} & \blu{81.88} & \blu{90.72} & \blu{96.53} & \blu{97.68} & \blu{98.64} & \blu{98.93} & \blu{99.38} \\
w/o Edge MPNN & 75.99 & 87.70 & 93.87 & 97.91 & 98.08 & 98.58 & 99.51 \\
MPNN→GAT & 33.19 & 66.36 & 71.88 & 80.82 & 79.85 & 83.81 & 95.73 \\
MPNN→GCN & 42.46 & 62.92 & 72.62 & 83.65 & 79.73 & 83.56 & 93.71 \\
Freeze encoder & 76.42 & 90.90 & 93.98 & 95.49 & 96.33 & 96.60 & 97.93 \\
\blu{Contrastive SSL} & \blu{72.31} & \blu{88.36} & \blu{95.20} & \blu{97.70} & \blu{97.60} & \blu{98.19} & \blu{99.44} \\
\midrule
Ours & \textbf{88.81} & \textbf{94.75} & \textbf{96.99} & \textbf{98.20} & \textbf{98.94} & \textbf{99.02} & \textbf{99.57} \\
\bottomrule
\end{tabular}
\caption{Ablation study on MFInstSeg dataset. We report Accuracy (\%) under varying ratios of labeled training data.
We compare our method with the method of removing the virtual node in gAAG (w/o VN), folding only once (Folding (1)) or three times (Folding (3)) in the geometric information decoder of our masked autoencoder, \blu{replacing the FoldingNet grid decoder with a plain MLP (FoldingNet $\rightarrow$ MLP),} removing edge MPNN in our masked autoencoder (w/o Edge MPNN),  replacing MPNN with other GNNs (MPNN→GAT and MPNN→GCN), and freezing the encoder of our autoencoder during fine-tuning stage. 
\blu{We also include a non-MAE self-supervised baseline based on graph contrastive learning (Contrastive SSL).}
Best results are in bold.}
\label{tab:ablation_acc}
\end{table}

\subsection{Ablation study}\label{sec:ablation}
To validate the effectiveness of components in BRepMAE, we conduct comprehensive ablation studies on the MFInstSeg dataset by removing or replacing these components and evaluating their impact on model performance across different data ratios (Tab.~\ref{tab:ablation_acc}).
%
%
Among all results, our algorithm performs best.
The specific analysis of the results is as follows:

\paragraph{Mask ratio}
\blu{We evaluate mask ratios from 0.1 to 0.9 under the same MFInstSeg ablation protocol as Tab.~\ref{tab:ablation_acc}. 
As shown in Tab.~\ref{tab:mask_ratio}, moderate-to-high masking ratios (0.7-0.9) generally achieve the best results, while too low masking (e.g., 0.1-0.3) tends to degrade performance, especially in the extremely low-label case. We use 0.8 as the default setting since it delivers consistently competitive accuracy across label ratios.}

\begin{table}[t!]
\centering
\scriptsize
\setlength{\tabcolsep}{1.5pt}
\begin{tabular}{lccccccc}
\toprule
\textbf{Mask ratio} & \textbf{0.1\%} & \textbf{0.5\%} & \textbf{1\%}  & \textbf{3\%} & \textbf{8\%} & \textbf{12.5\%} & \textbf{100\%} \\
\midrule
0.8 (default) & \textbf{88.81} & \textbf{94.75} & 96.99 & 98.20 & \textbf{98.94} & \textbf{99.02} & 99.57 \\
0.1 & 80.53 & 94.65 & 96.82 & 98.21 & 98.61 & 98.85 & 99.56 \\
0.2 & 82.95 & 93.62 & 96.93 & 98.45 & 98.64 & 98.86 & 99.56 \\
0.3 & 82.06 & 94.00 & 96.45 & 98.44 & 98.50 & 98.76 & 99.56 \\
0.4 & 87.14 & 94.68 & 96.11 & 97.96 & 98.42 & 98.85 & 99.57 \\
0.5 & 82.85 & 91.89 & 96.76 & 98.46 & 98.43 & 98.65 & 99.52 \\
0.6 & 83.34 & 93.97 & 96.86 & 98.46 & 98.57 & 98.88 & \textbf{99.58} \\
0.7 & 88.57 & 94.11 & 96.86 & \textbf{98.49} & 98.69 & 98.88 & 99.57 \\
0.9 & 87.89 & 94.11 & \textbf{97.07} & 98.02 & 98.58 & 98.90 & 99.56 \\
\bottomrule
\end{tabular}
\caption{\blu{Mask ratio sensitivity on MFInstSeg. 
We report Accuracy (\%) under varying ratios of labeled training data. 
Our default mask ratio is 0.8 and is used in all main experiments.}}
\label{tab:mask_ratio}
\end{table}

\paragraph{Loss weights}
\blu{We further study the sensitivity of the pre-training loss weights in Eq.~\eqref{equ:total-loss} by varying $(\alpha,\beta,\gamma,\delta)$ while keeping all other settings identical to Tab.~\ref{tab:ablation_acc}. Tab.~\ref{tab:loss_weight} shows that our method is generally stable across a wide range of weight choices, and emphasizing the geometric reconstruction term (larger $\beta$) tends to benefit the extremely low-label regime.}

\begin{table}[t!]
\centering
\scriptsize
\setlength{\tabcolsep}{1.0pt}
\begin{tabular}{@{}>{\raggedright\arraybackslash}p{0.30\linewidth}ccccccc@{}}
\toprule
\textbf{$(\alpha,\beta,\gamma,\delta)$} & \textbf{0.1\%} & \textbf{0.5\%} & \textbf{1\%}  & \textbf{3\%} & \textbf{8\%} & \textbf{12.5\%} & \textbf{100\%} \\
\midrule
Default (0.40, 0.36, 0.12, 0.12) & 88.81 & 94.75 & 96.99 & 98.20 & \textbf{98.94} & \textbf{99.02} & \textbf{99.57} \\
(0.25, 0.25, 0.25, 0.25) & 84.44 & 94.07 & 96.28 & 98.34 & 98.43 & 98.74 & \textbf{99.57} \\
(0.70, 0.10, 0.10, 0.10) & 81.95 & 90.56 & 95.69 & \textbf{98.41} & 98.56 & 98.91 & 99.50 \\
(0.10, 0.70, 0.10, 0.10) & \textbf{89.31} & \textbf{95.78} & \textbf{97.34} & 98.40 & 98.75 & 98.99 & 99.53 \\
\bottomrule
\end{tabular}
\caption{\blu{Loss-weight sensitivity on MFInstSeg. We report Accuracy (\%) under varying ratios of labeled training data with different $(\alpha,\beta,\gamma,\delta)$ in Eq.~\eqref{equ:total-loss}.}}
\label{tab:loss_weight}
\end{table}

\paragraph{UV sampling resolution}
\blu{We study the sensitivity to the UV-grid sampling resolution by re-running the full pipeline (preprocessing, pre-training, and fine-tuning) with different resolutions.}
\blu{Tab.~\ref{tab:uv_res} shows that the performance is generally stable across resolutions, with 15$\times$15 giving slightly higher accuracy in the extremely low-label regime.}
\blu{In our experiments, we use 10$\times$10 by default, which provides a good balance between accuracy and preprocessing/training cost.}

\begin{table}[t]
\centering
\scriptsize
\setlength{\tabcolsep}{1.0pt}
\begin{tabular}{@{}>{\raggedright\arraybackslash}p{0.30\linewidth}ccccccc@{}}
\toprule
\textbf{Resolution} & \textbf{0.1\%} & \textbf{0.5\%} & \textbf{1\%}  & \textbf{3\%} & \textbf{8\%} & \textbf{12.5\%} & \textbf{100\%} \\
\midrule
5$\times$5 & 85.81 & 94.62 & 96.88 & 97.79 & 98.53 & 98.71 & 99.58 \\
10$\times$10 (default) & 88.81 & 94.75 & 96.99 & 98.20 & \textbf{98.94} & \textbf{99.02} & 99.57 \\
15$\times$15 & \textbf{89.86} & \textbf{95.28} & \textbf{97.41} & \textbf{98.60} & 98.72 & \textbf{99.02} & \textbf{99.59} \\
30$\times$30 & 88.34 & 92.92 & 96.89 & 98.53 & 98.68 & 98.95 & 99.57 \\
\bottomrule
\end{tabular}
\caption{\blu{UV-grid resolution sensitivity on MFInstSeg. We report Accuracy (\%) under varying ratios of labeled training data. For a resolution $K{\times}K$, the number of samples along the face $u$ direction and the face $v$ direction is set to $K$, and the number of samples along the edge is set to $K$.}} 
\label{tab:uv_res}
\end{table}

\paragraph{Linear probing and frozen-encoder transfer}
\blu{We evaluate both a linear head and a 2-layer MLP head with either a frozen encoder (probing) or end-to-end fine-tuning under the same MFInstSeg protocol. The results in Tab.~\ref{tab:probing} show that the pre-trained encoder yields transferable representations, and that end-to-end fine-tuning further improves performance.}

\begin{table}[t!]
\centering
\scriptsize
\setlength{\tabcolsep}{1.0pt}
\begin{tabular}{@{}>{\raggedright\arraybackslash}p{0.30\linewidth}ccccccc@{}}
\toprule
\textbf{Head / Fine-tuning} & \textbf{0.1\%} & \textbf{0.5\%} & \textbf{1\%}  & \textbf{3\%} & \textbf{8\%} & \textbf{12.5\%} & \textbf{100\%} \\
\midrule
Linear probing (frozen) & 75.88 & 82.02 & 85.92 & 87.75 & 86.66 & 87.58 & 90.74 \\
Linear head (end-to-end) & 84.81 & 95.19 & 96.81 & 98.35 & 98.64 & 98.94 & 99.52 \\
2-layer MLP probing (frozen) & 76.94 & 84.30 & 88.91 & 92.29 & 91.88 & 93.34 & 96.55 \\
2-layer MLP (end-to-end) & 86.86 & 95.20 & 96.92 & 98.42 & 98.56 & 98.91 & 99.51 \\
\bottomrule
\end{tabular}
\caption{\blu{Linear probing and frozen-encoder transfer on MFInstSeg. We report Accuracy under various ratios of labeled training data.}}
\label{tab:probing}
\end{table}

\paragraph{Virtual Node}
\blu{Removing the virtual node from our gAAG representing the input BRep model leads to a significant performance drop, particularly in low-data cases.
Accuracy decreases by 6.47\% at a 0.1\% data ratio, indicating that introducing a global node to aggregate graph-level context is beneficial for downstream recognition.}

\paragraph{Geometric information decoder}
The FoldingNet geometric information decoder shows clear advantages over simpler MLP alternatives. 
\blu{Replacing the FoldingNet grid decoder with a plain MLP results in a consistent performance drop, most notably in the low-label regime (e.g., 0.1\%: 81.88\% vs.\ 88.81\%). This shows that using a stronger geometric decoder is beneficial, but a weaker decoder does not erase the advantage of BRepMAE, implying that the gain mainly comes from the pre-trained encoder representations.}
Moreover, our method performs the folding stage twice for reconstruction. 
In contrast, folding only once for reconstruction results in substantial performance degradation across all data ratios, with the largest drop of 5.71\% at a 1\% data ratio. 
Folding three times performs better but still underperforms our model, confirming that our choice produces the best.

\paragraph{Edge MPNN}
The alternating node-edge update mechanism in MPNN is critical for model performance. 
We remove the edge update step (w/o Edge MPNN) to test its impact on model performance.
The results show that it causes a significant performance decline, especially in few-shot scenarios (12.81\% drop at 0.1\% data).
This shows that edges play an important role in message passing, and not updating edge features can affect the transmission of node information, thereby reducing the model's performance.

\paragraph{MPNN vs. Alternative GNNs}
We also replace our MPNN module with GAT~\cite{velivckovic2018graph} or GCN~\cite{kipf2017semi} for comparisons.
The results of GAT show the worst performance with only 33.19\% accuracy at 0.1\% data, while the results of GCN perform slightly better but are still significantly worse than ours. 
The performance degradation stems from the fact that standard GCN and GAT models do not use edge features to update node features. 
In the BRep model, edges not only represent the adjacent relationships between faces (topological information) but also encapsulate the trimming information of the associated faces (geometric information). 
Thus, edge features play a crucial role in our representation learning task.

\paragraph{Contrastive self-supervised learning (SSL)}
\blu{We include a graph contrastive-learning baseline~\cite{Zhu:2020vf} to verify whether the MAE formulation is essential. 
As shown in Tab.~\ref{tab:ablation_acc}, this non-MAE baseline consistently underperforms our BRepMAE, with a large gap in the extremely low-label cases (e.g., 0.1\%: 72.31\% vs.\ 88.81\%), suggesting that masked reconstruction is more effective for learning transferable B-Rep representations.}

\paragraph{Freeze encoder during fine-tuning} 
We further study the impact of freezing the pre-trained encoder and training only the classification head. 
Compared with our full end-to-end fine-tuning, this setting yields consistently worse performance across all data ratios. 
For example, at the 0.1\% data ratio, the accuracy drops from 88.81\% to 76.42\% ($-$12.39\%). 
Even with abundant labels, the gap remains notable (e.g., 3\%: 95.49\% vs 98.20\%; 100\%: 97.93\% vs 99.57\%). 
These results indicate that while the pre-trained encoder provides strong, generic representations, adapting it to the target label space and dataset distribution through end-to-end fine-tuning is crucial for achieving optimal performance.

Overall, the ablation study confirms that each component of our BRepMAE framework makes a meaningful contribution to overall performance, with the MPNN encoder and geometric information decoder being the most critical for effective B-Rep representation learning.

\section{Conclusion and Discussion}\label{sec:conclusion}
This paper proposes a self-supervised learning framework based on the BRep model for MFR.
By utilizing self-supervised learning, our model can learn the representation of the BRep model from unlabeled data, enabling us to achieve model transfer with a small amount of labeled data.
%
%
%
%
%
We compared our algorithm with two other supervised models and a point cloud-based self-supervised model, demonstrating that it achieves state-of-the-art recognition accuracy and generalization.
In short, our method can learn generalizable representations from unlabeled B-Rep data that can be applied to downstream tasks after a brief fine-tuning stage.
This ability is of paramount importance for practical industrial applications, since acquiring large-scale, high-quality labeled datasets is often prohibitively expensive.

\paragraph{Recognition accuracy}
%
%
Despite strong overall performance, our method’s recognition is not 100\% accurate, leading to occasional missegmentation, particularly for geometrically similar or topologically complex features. This may stem from subtle differences in high-dimensional feature representations or limited training samples for rare feature types.
In the future, we plan to increase the diversity and quantity of training data, thereby enhancing the model's ability to recognize rare or complex features and further improve its stability and robustness.
\blu{Moreover, the virtual node aggregates the global context of each gAAG, and its embedding can be used directly as a graph-level representation (e.g., for CAD model retrieval), which we leave for future exploration.}

\paragraph{Process information}
%
Our current approach focuses on geometric and topological information and does not incorporate critical process-related information, such as manufacturing precision and surface roughness, which are essential for practical application as they directly guide strategy and parameter selection.
%
%
%
In the future, we plan to extend our dataset and model with process information to enable adaptation to downstream tasks such as process planning, parameter optimization, and manufacturability analysis, thereby enhancing their practical value in real industrial scenarios.

\paragraph{Intersecting features}
Feature intersection occurs when multiple features overlap, creating faces that belong to more than one machining feature, which challenges the proposed single-label segmentation methods. This limitation is especially evident in complex industrial parts with common feature fusion and intricate topologies.
%
%
To address this issue, future research will explore multi-label segmentation strategies, supported by more expressive network architectures, refined labeling schemes, and expanded training datasets with richer examples of feature intersection.

\bibliography{src/reference}

\appendix
\section{Details of our encoders}
Tables~\ref{tab:face-encoder} and~\ref{tab:edge-encoder} show the details of our encoders.

\begin{table}[!t]
\footnotesize
\setlength{\lightrulewidth}{0.3pt}
\begin{tabularx}{0.48\textwidth}{p{2.8cm}p{2.7cm}p{2.7cm}}
    \hline
    \textbf{Operator} & \textbf{Input} & \textbf{Output}  \\
    \midrule
    Conv2D, $k = 3$ & $7\times 10 \times 10$ & $32\times 10 \times 10$ \\
    BatchNorm2D & $32\times 10 \times 10$ & $32\times 10 \times 10$\\
    Mish & $32\times 10 \times 10$ & $32\times 10 \times 10$\\
    Conv2D, $k = 3$ & $32\times 10 \times 10$ & $64\times 10 \times 10$ \\
    BatchNorm2D & $64\times 10 \times 10$ & $64\times 10 \times 10$\\
    Mish & $64\times 10 \times 10$ & $64\times 10 \times 10$\\
    Conv2D, $k = 3$ & $64\times 10 \times 10$ & $128\times 10 \times 10$ \\
    BatchNorm2D & $128\times 10 \times 10$ & $128\times 10 \times 10$\\
    Mish & $128\times 10 \times 10$ & $128\times 10 \times 10$\\
    AdaptiveAvgPool2D & $128\times 10 \times 10$ & $128\times 1 \times 1$\\
    Flatten & $128\times 1 \times 1$ & $128$\\
    \hline
\end{tabularx}
\caption{Details of our face encoder, the convolution kernel size is denoted as $k$.
The size of the grids is $10\times 10$.
}
\label{tab:face-encoder}
\centering
\end{table}

\begin{table}[!t]
\footnotesize
\setlength{\lightrulewidth}{0.3pt}
\begin{tabularx}{0.48\textwidth}{p{3.3cm}p{2.8cm}p{3.3cm}}
    \hline
    \textbf{Operator} & \textbf{Input} & \textbf{Output}  \\
    \midrule
    Conv1D, $k = 3$ & $12\times 10$ & $32\times 10$ \\
    BatchNorm1D & $32\times 10$ & $32\times 10$\\
    ReLU & $32\times 10$ & $32\times 10$\\
    Conv1D, $k = 3$ & $32\times 10$ & $64\times 10$ \\
    BatchNorm1D & $64\times 10 $ & $64\times 10$\\
    ReLU & $64\times 10$ & $64\times 10$\\
    AdaptiveAvgPool1D & $64\times 10$ & $64\times 1$\\
    Flatten & $64\times 1$ & $64$\\
    \hline
\end{tabularx}
\caption{Details of our edge encoder, the convolution kernel size is denoted as $k$.
We uniformly sample 10 points along the curve parameter domains.
}
\label{tab:edge-encoder}
\centering
\end{table}

\section{BRep Embedding}
Our BRep feature encoder consists of a 2D CNN for face geometry, a 1D CNN for edge geometry, and three MLPs for face bounding box, face attributes, and edge attributes.
%
Each UV grid of the BRep faces has a 7-dimensional feature, and each face has $10\times 10$ UV grids.
Thus, our face geometry encoder takes a shape of $7\times 10\times 10$ vector as input and outputs a 128-dimensional vector as the extracted information of faces.
Similarly, our edge geometry encoder takes the sampling points along the curve parameter domains with a shape of $12\times 10$ and generates a 64-dimensional vector to represent the edge.
The details of these two encoders are shown in Tab.~\ref{tab:face-encoder} and~\ref{tab:edge-encoder}. 
The input vector of the face bounding box encoder is 6-dimensional.
It is transformed into a 64-dimensional vector by an MLP with linear, ReLU, and normalization layers, followed by linear layers.
The output channel of the first linear layer is 128.
The structure of the face and edge attribute encoders is straightforward, consisting of only a linear layer and a normalization layer. 
These two MLPs convert the input vectors into 64-dimensional vectors.

The weights of these encoders are shared across all elements in a graph to ensure permutation invariance~\cite{jayaraman2021uv}.
Finally, we concatenate the above embeddings to construct the representation for each node and edge. 
For each node, the embeddings are fused into a 256-dimensional vector $\mathbf{x}_i$; for each edge, the corresponding embeddings are fused into a 128-dimensional vector and then through a linear layer to convert it into a 256-dimensional vector $\mathbf{e}_ {ij}$.
In addition to better leveraging global information, we introduce a virtual node to gAAG that is connected to all nodes, whose node embedding is also a 256-dimensional vector, and the edge embeddings are 128-dimensional vectors.
Therefore, the gAAG is defined as $\mathcal{G} = (\mathcal{V}, \mathcal{E}, \mathbf{X}, \mathbf{X_E})$, where $\mathcal{V}$ is the set of nodes containing the virtual node, $\mathcal{E}$ is the set of edges containing the edges between virtual node and other nodes, $\mathbf{X} \in \mathbb{R}^{|\mathcal{V}| \times 256}$ is the face embeddings and $\mathbf{X_E} \in \mathbb{R}^{|\mathcal{E}| \times 256}$ is the edge embeddings.

\section{Three steps of MPNN}
The key procedure in the MPNN is message passing, which has three steps: Message, Aggregate, and Update.
In the Message step, each node in the graph sends a message to its neighbors, computed from the features of the sending and receiving nodes and the edge between them.
Second, each node aggregates the messages it receives from its neighbors.
The aggregate function is usually sum, mean, max, or min to ensure it is permutation-invariant.
Lastly, each node updates its feature using a differentiable function, such as a linear layer or an activation function, from the aggregate messages and its original feature.

\section{Loss functions}
 \label{supp:loss}

\begin{table}[!t]
\footnotesize
\setlength{\lightrulewidth}{0.3pt}
\begin{tabularx}{0.48\textwidth}{p{1.3cm}p{2.1cm}p{2.2cm}p{2.3cm}}
    \hline
    \textbf{Stage} & \textbf{Operator} & \textbf{Input} & \textbf{Output}  \\
    \midrule
    \multirow{3}{*}{ \shortstack{   \begin{minipage}{1.5cm} 
    \raggedright
    First \\ folding
  \end{minipage}}} & Linear               & 256+2                       & 512                       \\
     & ReLU                 & 512                       & 512                       \\
    & Linear               & 512                       & 64                        \\
    \midrule 
    \multirow{3}{*}{\shortstack{   \begin{minipage}{1.5cm} 
    \raggedright
    Second \\ folding
  \end{minipage}}} & Linear               & 256+64                       & 512                       \\
    & ReLU                 & 512                       & 512                       \\
    & Linear               & 512                       & 7                         \\
    \hline
\end{tabularx}
\caption{Details of our geometric information decoder.
We perform two folding stages consisting of MLPs to decode the geometric information.
}
\label{tab:geom-decoder}
\centering
\end{table}

\begin{table}[!t]
\footnotesize
\setlength{\lightrulewidth}{0.3pt}
\begin{tabularx}{0.48\textwidth}{p{3cm}p{3cm}p{3cm}}
    \hline
    \textbf{Operator} & \textbf{Input} & \textbf{Output}  \\
    \midrule
        Linear            & 256                      & 128                       \\
        ReLU              & 128                      & 128                       \\
        Dropout (p=0.1)   & 128                      & 128                       \\
        Linear            & 128                      & 10                        \\
    \hline
\end{tabularx}
\caption{Attribute decoder used in pre-training to reconstruct face attributes from the latent node feature.
}
\label{tab:attr-decoder}
\centering
\end{table}

\begin{table}[!t]
\footnotesize
\setlength{\lightrulewidth}{0.3pt}
\begin{tabularx}{0.48\textwidth}{p{3cm}p{3cm}p{3cm}}
    \hline
    \textbf{Operator} & \textbf{Input} & \textbf{Output}  \\
    \midrule
        Linear            & 256                      & 64                        \\
        ReLU              & 64                       & 64                        \\
        Dropout (p=0.1)   & 64                       & 64                        \\
        Linear            & 64                       & 6                         \\
    \hline
\end{tabularx}
\caption{Bounding-box decoder used in pre-training to reconstruct face AABB parameters from the latent node feature.
}
\label{tab:aabb-decoder}
\centering
\end{table}

\paragraph{Feature term}
The feature term $\mathcal{L}_\text{feat}$ ensures that our autoencoder can correctly reconstruct the masked node features.
It is defined by the Mean Squared Error (MSE) as follows:
\begin{equation}\label{equ:feat-loss}
    \mathcal{L}_\text{feat} = \frac{1}{|\mathbf{X}_\text{m}|}\sum_{\mathbf{x}_i\in \mathbf{X}_\text{m}, \hat{\mathbf{x}}_i\in \hat{\mathbf{X}}_\text{m}}\|\mathbf{x}_i - \hat{\mathbf{x}}_i \|_2^2,
\end{equation}
where $\hat{\mathbf{X}}_\text{m}$ is the set of reconstructed features of the masked node in $\mathcal{G}$, $\mathbf{X}_\text{m}$ is the ground truth, and $|\mathbf{X}_\text{m}|$ is the number of features in $\mathbf{X}_\text{m}$.

\paragraph{Attribute information term}
The predicted surface attribute information $\hat{f}_{\text{attr}, i}$ is produced by an attribute decoder from the node feature $\mathbf{x}_i$ it corresponds to.
The architecture of the attribute decoder is shown in Tab.~\ref{tab:attr-decoder}.
We apply the following function defined by the MSE to ensure the predicted $\hat{f}_{\text{attr}, i}$ is close to the ground truth $f_{\text{attr}, i}$:
\begin{equation}\label{equ:attr-loss}
    \mathcal{L}_\text{attr} =  \frac{1}{|\mathbf{X}_\text{m}|}\sum_{i=0}^{|\mathbf{X}_\text{m}|-1}\| f_{\text{attr}, i} - \hat{f}_{\text{attr}, i} \|_2^2,
\end{equation}

\paragraph{Bounding box term}
The same as the attribute information term, we propose a bounding box decoder to predict $f_\text{aabb}$ from $\textbf{x}_i$.
Tab.~\ref{tab:aabb-decoder} shows the architecture of the bounding box decoder.
The bounding box term is defined as follows:
\begin{equation}\label{equ:aabb-loss}
    \mathcal{L}_\text{aabb} =  \frac{1}{|\mathbf{X}_\text{m}|}\sum_{i=0}^{|\mathbf{X}_\text{m}|-1}\| f_{\text{aabb}, i} - \hat{f}_{\text{aabb}, i} \|_2^2,
\end{equation}
where $\hat{f}_{\text{aabb}, i}$ is the predicted bounding box information and $f_{\text{aabb}, i}$ is the ground truth.

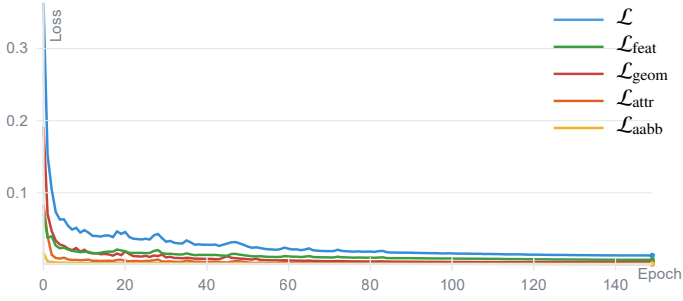
\begin{figure}[t]
 \centering
		\begin{overpic}[width=0.99\linewidth]{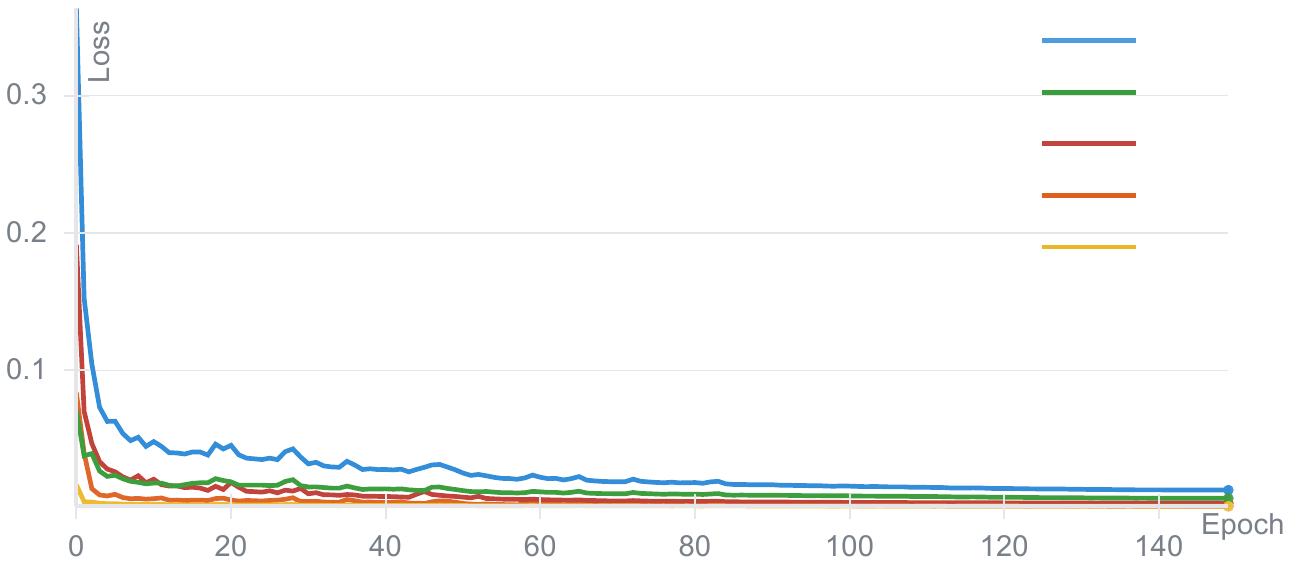}
			{
			    \put(90,40){\footnotesize $\mathcal{L}$}
                    \put(90,36){\footnotesize $\mathcal{L}_\text{feat}$}
                    \put(90,32){\footnotesize $\mathcal{L}_\text{geom}$}
			    \put(90,28){\footnotesize $\mathcal{L}_\text{attr}$}
                    \put(90,24){\footnotesize $\mathcal{L}_\text{aabb}$}
		    }
		\end{overpic}
		 \vspace{-3mm}
		 \caption{
         The loss curves of our model during the pre-training stage. After 150 epochs, our model has basically converged.
			 }
		 \label{fig:loss-curve}
\end{figure}

\paragraph{Loss curves}
The loss curves in the pre-training stage are illustrated in Fig.~\ref{fig:loss-curve}.

\section{Architecture of the classification head used for fine-tuning}
This MLP (Table~\ref{tab:clas-head}) processes the node embeddings (256-dimensional) by passing them through a sequence of linear layers with hidden dimensions of 1024 and 256. 
Each linear layer is followed by a ReLU activation and a dropout layer with a rate of 0.5 for regularization. 
The final layer maps the 256-dimensional features to a logit vector of size $n_c$, where $n_c$ is the total number of predefined machining feature classes (e.g., hole, slot, pocket, fillet, etc.).

\begin{table}[!t]

\footnotesize
\setlength{\lightrulewidth}{0.3pt}
\begin{tabularx}{0.48\textwidth}{p{3cm}p{3cm}p{3cm}}
    \hline
    \textbf{Operator} & \textbf{Input} & \textbf{Output}  \\
    \midrule
        Linear            & 256                      & 1024                        \\
        ReLU              & 1024                       & 1024                        \\
        Dropout (p=0.5)   & 1024                       & 1024                        \\
        Linear            & 1024                       & 256                         \\
        ReLU              & 256                       & 256                        \\
        Dropout (p=0.5)   & 256                       & 256                        \\
        Linear            & 256                       & 25                      \\
    \hline
\end{tabularx}
\caption{Architecture of the classification head used for fine-tuning. 
}
\label{tab:clas-head}
\centering
\end{table}

\section{Five common failure patterns}
\begin{figure*}[t]
 \centering
 \includegraphics[width=0.95\textwidth]{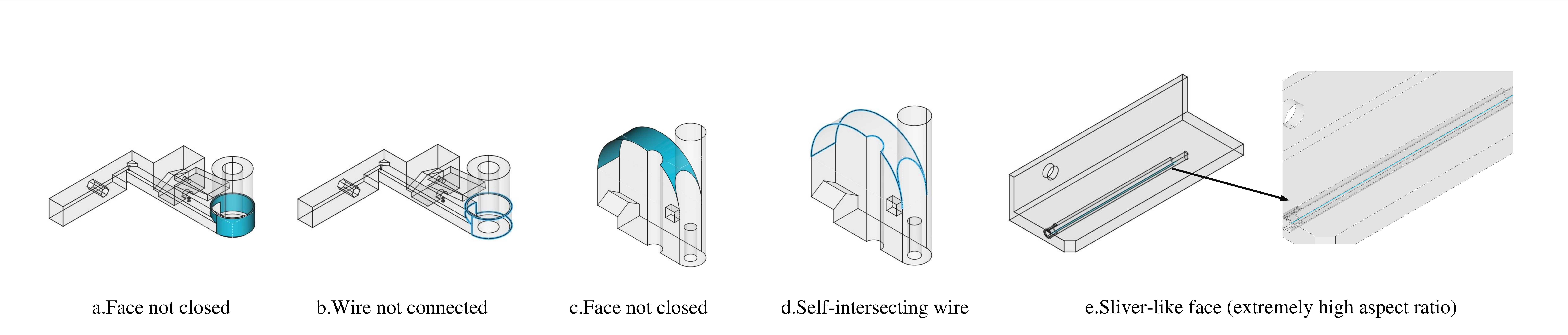}
 \vspace{-2mm}
 \caption{Typical invalid geometries encountered during preprocessing. Gray denotes the original CAD geometry, and blue highlights the abnormal sub-shapes reported by CAD-kernel geometry checks.}
\label{fig:Invalid}
\end{figure*}

Fig.~\ref{fig:Invalid} illustrates five common failure patterns: (a,c) face not closed (the face boundary cannot form a valid closed loop, so the face is not a valid trimming region); (b) wire not connected (edge segments in a loop are not properly connected at endpoints); (d) self-intersecting wire (the loop self-intersects/overlaps in the parameter domain or its projection, hence it is not a simple closed curve); and (e) sliver-like face (extremely high aspect ratio), where a very small local thickness leads to numerically unstable sampling and geometric evaluation (e.g., normals/UV grids). Such cases may cause UV-grid extraction failures or unstable features and are filtered out during preprocessing.

\section{Metric definition}
Acc measures the proportion of correctly classified faces across all classes, providing a global performance assessment, and is defined as follows: 
The accuracy is defined as:
\begin{equation}
\text{Acc} = \frac{N_\text{cor}}{N_\text{total}},
\end{equation}
where $N_\text{cor}$ is the number of correctly classified faces, and $N_\text{total}$ is the number of total faces.

mIoU, calculated as the mean of IoU values across all machining feature classes, offers a more balanced evaluation that accounts for class imbalance and penalizes both false positives and false negatives. 
It measures the similarity and diversity of two sets of samples, and is defined as the ratio of the intersection to the union of the sample sets:
\begin{equation}
\text{mIoU} = \frac{1}{N_\text{c}} \sum_{c=0}^{N_\text{c}-1} \frac{A_c \cap B_c}{A_c \cup B_c},
\end{equation}
where $N_\text{c}$ represents the total number of classes, $A_c$ represents the predicted set of faces for class $c$, and $B_c$ represents the ground truth set of faces for class $c$.

\section{Comparisons}

\subsection{Comparisons with full-supervised methods}
\paragraph{Comparisons on MFCAD2}
When only 0.1\% of the labeled faces are available, our framework delivers 75.99\% accuracy, whereas AAGNet and BRepMFR reach only 51.22\% and 26.53\%, respectively. 
With 1\% labeled data, our accuracy already reaches 96.25\%, demonstrating high utilization of labeled data.
A similar trend can be observed for the mIoU metric. 
At the 0.1\% ratio, BRepMAE achieves 49.33\% mIoU compared to 38.23\% for AAGNet and 6.01\% for BRepMFR. 
Beyond 2\% data, our method surpasses 94\% mIoU and eventually saturates at 98.87\% with the full training set. 

\paragraph{Comparisons on CADSynth}

With only 0.1\% of the labels, BRepMAE achieves 93.89\% accuracy, surpassing AAGNet (58.53\%) and BRepMFR (31.97\%). 
In addition, the full-supervised methods lag behind our 0.1\% result with 10 times as many labels (1\%) provided, which further proves the high utilization rate of our algorithm for labeled data.
%
The mIoU results exhibit the same pattern: BRepMAE achieves 79.34\% mIoU at the 0.1\% ratio, while AAGNet and BRepMFR reach only 47.17\% and 3.42\%, respectively. 
Across the remaining ratios, our method maintains at least a few percentage points lead and ultimately reaches 99.33\% mIoU with full training data. 

\paragraph{Comparisons on MFInstSeg}
When the labeled ratio is 0.1\%, BRepMAE already reaches 81.49\% accuracy, compared to 46.48\% for AAGNet and 20.83\% for BRepMFR. 
Increasing the ratio to 1\% boosts our accuracy to 97.27\%, illustrating our model adapts quickly once a handful of annotations is available. 
In terms of mIoU, BRepMAE improves from 63.27\% at 0.1\% data to 90.80\% at 1\%,  ahead of the baselines (34.01\%/9.66\% for AAGNet/BRepMFR at 0.1\%). 
With fully labeled data, our method maintains a small but consistent advantage (99.53\% accuracy and 98.42\% mIoU), confirming that our learnt representation benefits both low-labeled-data and high-labeled-data situations.

\section{Additional reconstruction results}
\begin{figure*}[!h]
 \centering
 \includegraphics[width=0.99\linewidth]{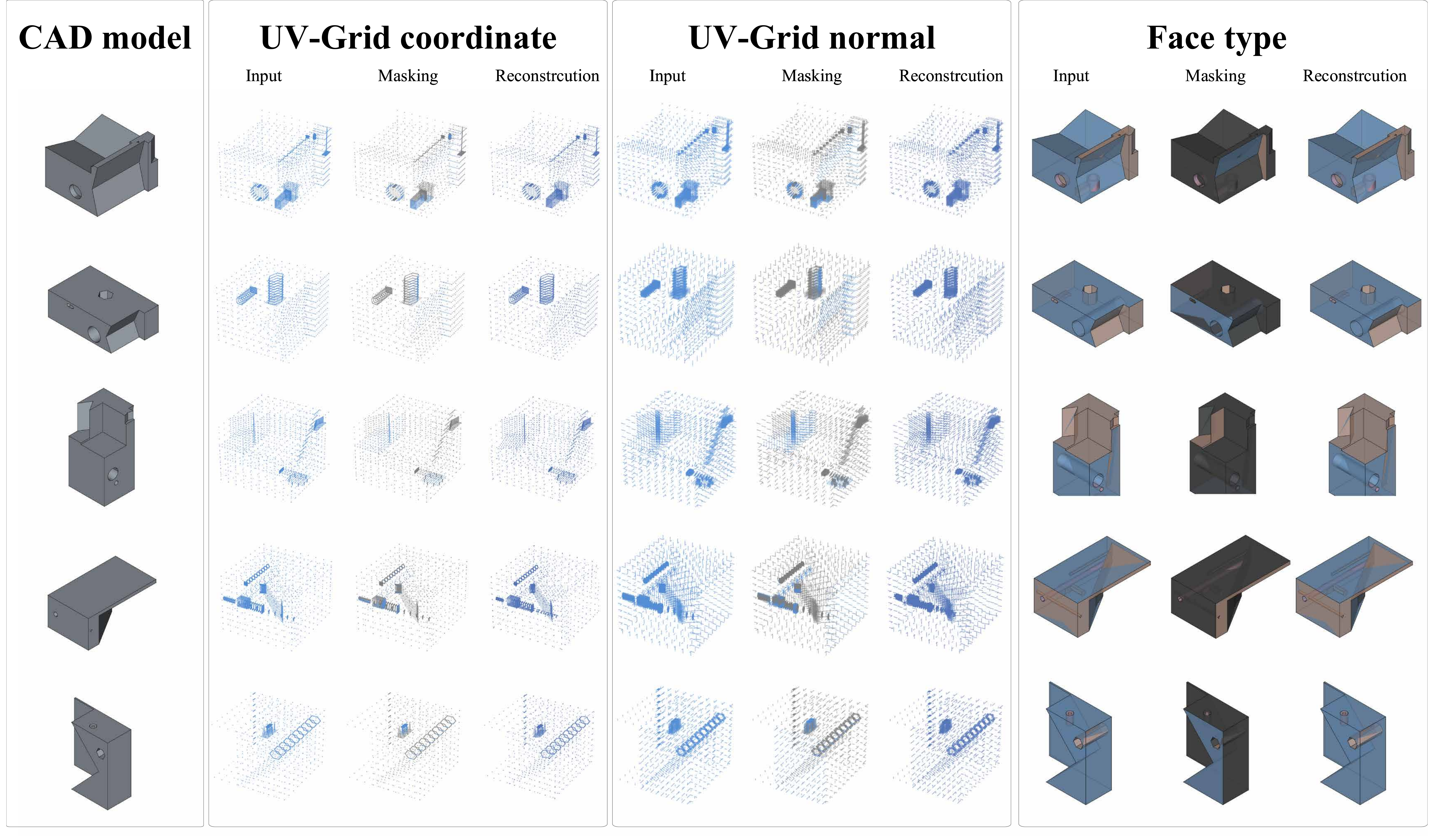}
 \caption{Additional reconstruction results. We set the masked parts of the CAD models to black; different face types are shown in different colors.}
 \label{fig:reconstruction-supp}
\end{figure*}

We provide additional qualitative reconstruction examples generated by our pre-training model in Fig.~\ref{fig:reconstruction-supp}.

\section{Additional machining feature recognition results}

\begin{figure*}[!h]
 \centering
		\begin{overpic}[width=0.95\linewidth]{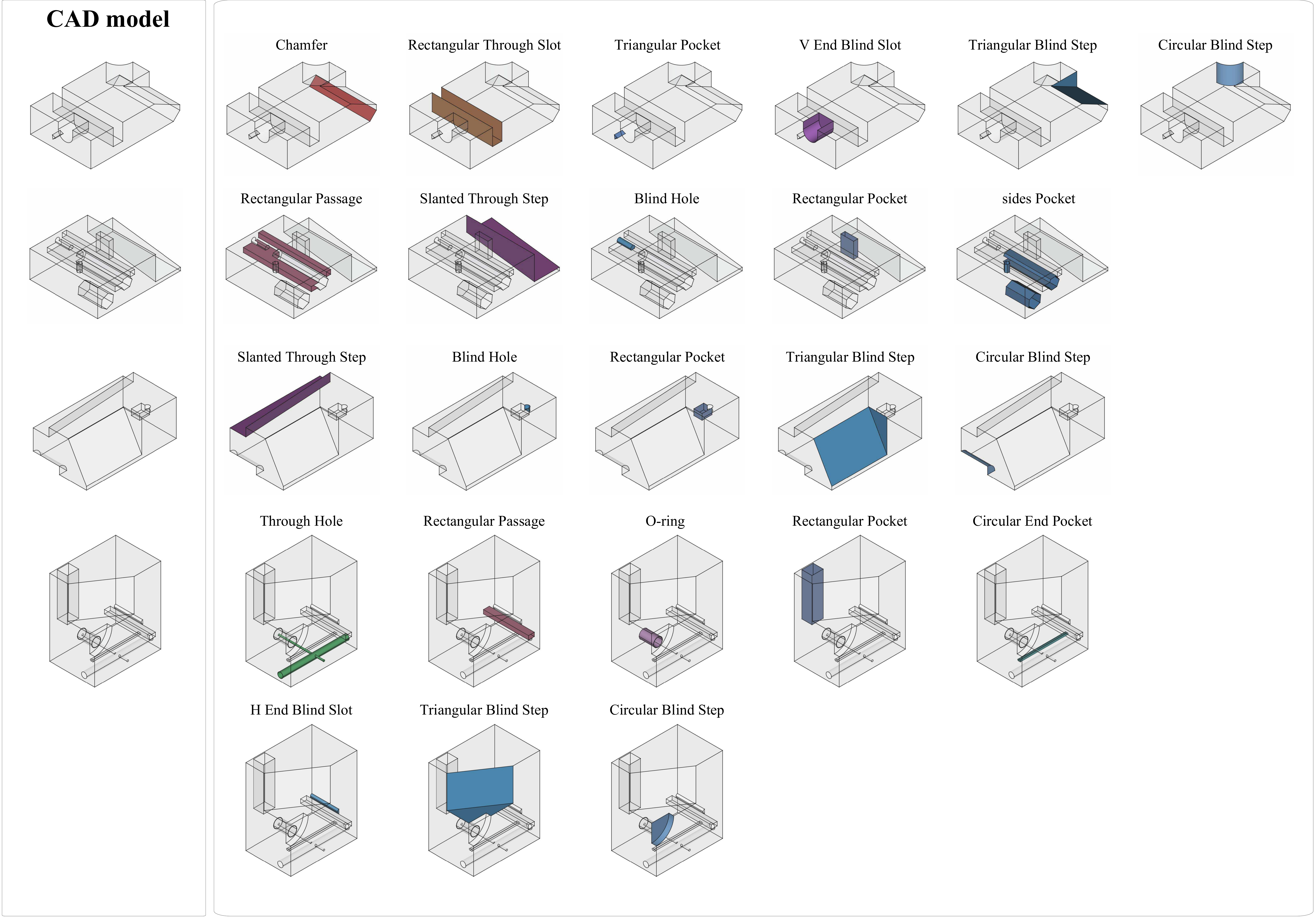}
			{
		    }
		\end{overpic}
		 \vspace{-3mm}
		 \caption{
         The machining feature recognition results.
         Different machining features are painted with different colors.
			 }
		 \label{fig:machining-feature}
\end{figure*}

We present the additional machining feature recognition results in Fig.~\ref{fig:machining-feature}.
The selected parts are from the validation set and cover a wide range of typical machining feature types, including chamfers, through slots, blind holes, steps, and various pockets. 
Our proposed BRepMAE model is employed to segment and classify the features on each B-Rep model.

As shown in the figure, the proposed method effectively segments and identifies diverse machining features in complex parts. 
Even in cases where multiple features intersect or overlap, resulting in significant changes to the topology and geometry, the model is able to distinguish and recognize each feature accurately. 
For example, in Part 2 and Part 3, the model successfully separates and classifies rectangular passages, blind holes, and pockets. 
For parts with more complex structures, such as slanted steps, O-rings, and various blind steps (e.g., Part 3 and Part 4), the method still demonstrates robust segmentation and recognition performance.

%
%

In summary, the proposed BRepMAE method demonstrates excellent performance in recognizing a wide variety of machining features in complex parts. 
It is well-suited for practical engineering applications involving diverse geometric and topological structures. 

\end{document}